\documentclass[11pt,a4paper]{article}
\pdfoutput=1
\usepackage{geometry}
\usepackage{graphicx}
\usepackage{color}
\usepackage{framed}
\usepackage{amssymb}
\usepackage{latexsym}
\usepackage{amsmath}
\usepackage{bm}
\usepackage[width=0.9\textwidth]{caption}
\usepackage[retainorgcmds]{IEEEtrantools}
\usepackage[bookmarks=true,pdfborder={0 0 0}]{hyperref}
\usepackage[all]{hypcap}
\title{Periodic Monopoles from Spectral Curves}
\author{Rafael Maldonado\footnote{{\tt rafael.maldonado@durham.ac.uk}}\vspace{0.2cm}\\\emph{Department of Mathematical 
Sciences,}\\\emph{South Road, Durham DH1 3LE, UK}}

\begin{document}
\maketitle
\begin{abstract}
\noindent We consider SU(2) Bogomolny equations on $\mathbb{R}^2\times\hat{S}^1$ and use the spectral curve 
defined by the holonomy in the periodic direction to approximate the fields in the limit of large size to 
period ratio.  Symmetries of the Nahm transform allow a study of the effective two dimensional dynamics, 
which is compared with known results on the full moduli space.  The techniques are applied to systems of 
higher charge and higher rank gauge group, allowing a direct comparison to other periodic Yang-Mills systems.
\end{abstract}
\section{Introduction}\label{introduction}
The Bogomolny equations on $\mathbb{R}^2\times\hat{S}^1$ (which we refer to as describing a `periodic 
monopole') were first introduced by Cherkis \& Kapustin \cite{CK01,CK02,CK03}.  Approximate analytical and 
numerical solutions of topological charge $1$ and $2$ were constructed by Harland and Ward \cite{War05,HW09} 
using the Nahm transform (see \cite{Jar04} for a review).  The remainder of this section describes the setup, 
which is illustrated by reference to the charge $1$ example of \cite{War05} in section \ref{spap}, where it 
is also shown that the monopole fields can be approximated as being two dimensional and Abelian.  Sections 
\ref{ch2spap} and \ref{ch2} apply these methods to the charge $2$ periodic monopole, previously studied in 
\cite{HW09}.  This will allow a study of various geodesics on the moduli space, whose asymptotic form agrees 
with that of \cite{CK02}.  We will also consider the relevant three dimensional dynamics and motion of 
`lumps' on the dual cylinder.  In section \ref{su3pdcmon} we look at several SU(3) configurations of low 
charge, while section \ref{singularities} shows explicitly the relation with the doubly periodic instanton 
\cite{FP+}.  The discussion is concluded with some ideas for future work in section \ref{outlook}.
\subsection{Monopole Data}\label{monopoledata}
BPS monopoles are described by a dimensional reduction of the self-dual Yang-Mills equations to three 
dimensions, such that the component of the gauge potential in the suppressed direction becomes a scalar Higgs 
field valued in the Lie algebra $\mathfrak{su}(N)$,
\begin{equation}
\hat{F}\,=\,\ast\hat{D}\hat{\Phi}.\label{bog}
\end{equation}
We will use coordinates $\rho\text{e}^{\text{i}\theta}=\zeta\in\mathbb{C}\cong\mathbb{R}^2$ and 
$z\in\mathbb{R}/\beta\mathbb{Z}$ and look for solutions periodic in one of the remaining spatial directions.  
The boundary conditions at large $\rho$ are chosen to match those of an Abelian chain, such that 
$\hat{\Phi}_\infty$ behaves as $\log(\rho)$ and the Bogomolny equations \eqref{bog} require 
$\hat{\Phi}_\infty$ to be a harmonic function on $\mathbb{R}^2\times\hat{S}^1$.  Imposing strict periodicity 
in $\theta$ and $z$ then requires $\theta$ dependence to enter $\hat{\Phi}_\infty$ at 
$\mathcal{O}(\rho^{-1})$ and $z$ dependence to contribute at $\mathcal{O}(\rho^{-1/2}\text{e}^{-\rho})$, well 
within the core region.
\par
\begin{figure}
\centering
\includegraphics[width=6cm]{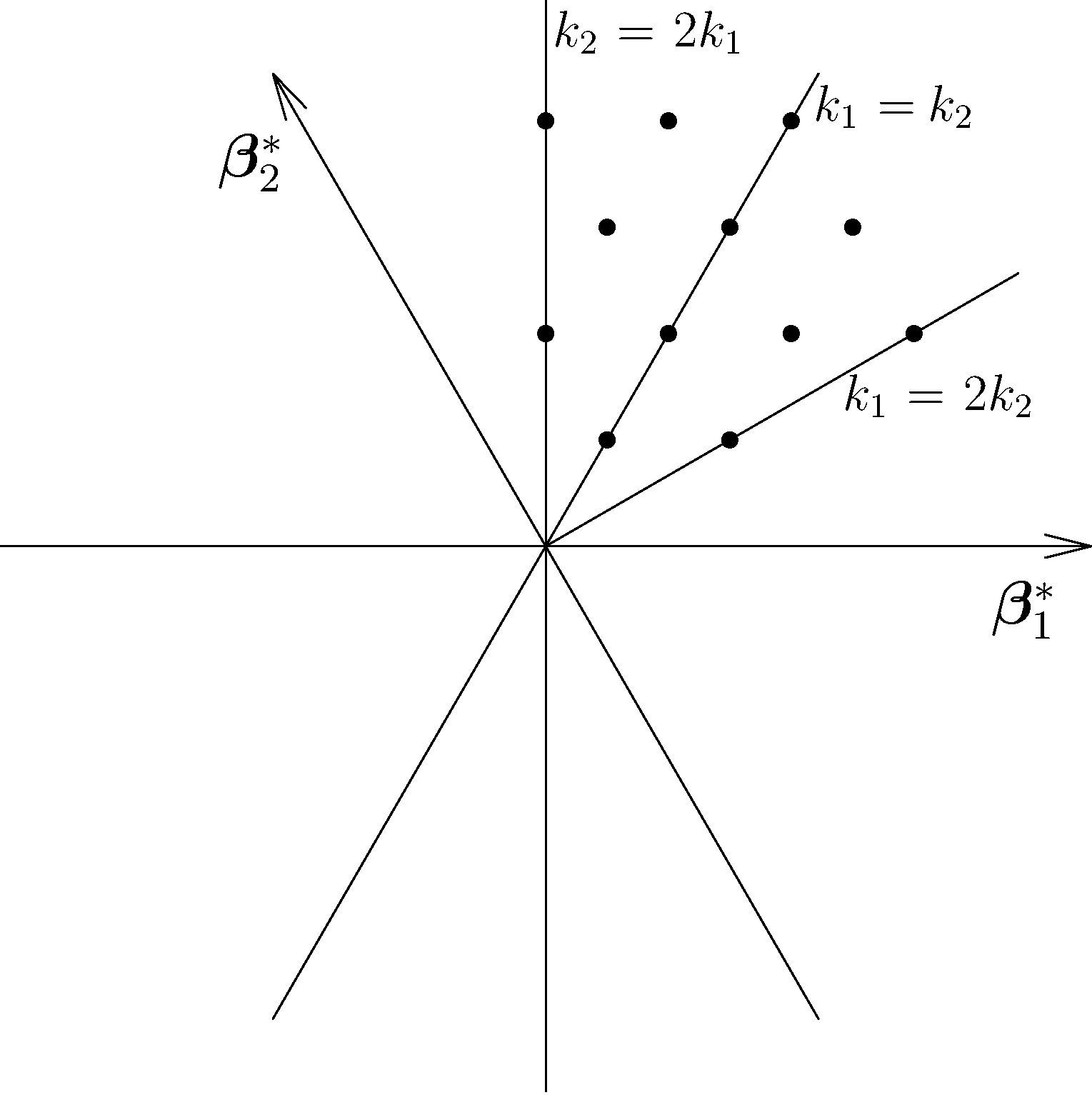}
\caption{Root diagram for SU(3) showing magnetic weights allowed by \protect\eqref{lcond}.}\label{fig01}
\end{figure}

 An SU($N$) monopole has boundary data defined by an $N$-component vector of integers, $\bm{\ell}$.  Recalling 
that the monopole fields are valued in $\mathfrak{su}(N)$ and noting that we are free to permute the entries 
in $\hat{\Phi}$ by a choice of gauge, the elements of $\bm{\ell}$ satisfy
\begin{equation}
\sum_{i=1}^N\ell_i\,=\,0\qquad\text{and}\qquad\ell_i\,\geq\,\ell_{i+1}.\label{lcond}
\end{equation}
We also have complex vectors $\bm{v}$, $\bm{b}$ and $\bm{\mu}$, whose components again sum to zero.  The 
asymptotic fields are then
\begin{IEEEeqnarray*}{rcl}
-\text{i}\beta\hat{\Phi}_\infty\,&=&\,\bm{\ell}\log(\rho)+\bm{v}+\Re(\bm{\mu}\zeta^{-1})+\mathcal{O}(\rho^{-2})\\
\text{i}\beta\hat{A}_{z,\infty}\,&=&\,\bm{\ell}\theta+\bm{b}+\Im(\bm{\mu}\zeta^{-1})+\mathcal{O}(\rho^{-2}),
\end{IEEEeqnarray*}
and are combined, defining $\bm{v}+\text{i}\bm{b}=\bm{\mathfrak{v}}$, into
\begin{equation}
\beta\hat{\phi}_\infty\,=\,-\text{i}\beta(\hat{\Phi}-\text{i}\hat{A}_z)_\infty\,=\,\bm{\ell}\log(\zeta)+\bm{\mathfrak{v}}+\bm{\mu}\zeta^{-1}+\mathcal{O}(\rho^{-2}).\label{asympfields}
\end{equation}
Such a monopole can be constructed by a minimal embedding of fundamental SU(2) monopoles in the 
$(N-1)$-dimensional co-root space with integer magnetic weights $k_i$ arranged into a vector $\bm{k}$,
\begin{equation*}
\bm{\ell}\,=\,\sum_{i=1}^N\ell_i\bm{e}_i\,=\,\sum_{i=1}^{N-1}k_i\bm{\beta}^\ast_i
\end{equation*}
where it is convenient to represent the co-root vectors in terms of $N$-dimensional vectors 
$\bm{\beta}^\ast_i=\bm{e}_i-\bm{e}_{i+1}$ and the $\{\bm{e}_i\}$ are basis vectors for $\bm{\ell}$.  The 
SU(3) case is illustrated in fig.~\ref{fig01}.
\par The monopole charge is given by the first Chern class,
\begin{equation}
q\,=\,\lim_{R\to\infty}\int_{\rho=R}\frac{\text{tr}(\hat{F}\hat{\Phi})}{4\pi\|\hat{\Phi}\|}\label{charge}
\end{equation}
where integration is over the 2-torus at radial infinity, the length of the Higgs field is 
$\|\hat{\Phi}\|^2=-\frac{1}{2}\text{tr}(\hat{\Phi}^2)$ and $\text{tr}(\cdot)$ denotes the trace in the Lie 
algebra.  For the $i^{\text{th}}$ fundamental monopole this evaluates to $q=k_i$.  It is possible to convert 
between the elements of $\bm{\ell}$ and those of $\bm{k}$ using
\begin{equation*}
k_j\,=\,\sum_{i=1}^{j}\ell_i\qquad\text{and}\qquad\ell_i\,=\,k_i-k_{i-1},
\end{equation*}
and we define $K=\text{max}(\{k_i\})$.  We will often refer to a specific class of $\text{SU}(N)$ monopole 
simply by its $(N-1)$-dimensional charge vector $\bm{k}$.
\par As is done for monopoles in $\mathbb{R}^3$ \cite{Har08,MS07,War82}, fundamental monopole masses are 
defined by the pattern of symmetry breaking of the leading terms in $\hat{\Phi}$.  In particular, the 
$i^{\text{th}}$ mass is
\begin{equation*}
m_i\,=\,\ell_i-\ell_{i+1}
\end{equation*}
where an interpretation as a physical mass requires the specification of a radial cut off.  If all the masses 
are non-zero, the SU($N$) symmetry is maximally broken by the asymptotic Higgs field to $\text{U}(1)^{N-1}$.  
Otherwise, there may be unbroken subgroups according to whether the corresponding $\mathfrak{v}_i$ are the 
same.  We will see examples of this in section \ref{minsb}.
\par Applying \eqref{charge} to the $\mathfrak{su}(N)$-valued fields, the total charge, $q$, is given by the 
product of fundamental charges and masses,
\begin{equation*}
q^2\,\propto\,\sum_{i=1}^N\ell_i^2\,=\,\sum_{i=1}^{N-1}k_im_i.
\end{equation*}
A similar result holds for SU($N$) monopoles in $\mathbb{R}^3$, although it is noteworthy that in contrast 
to the $\mathbb{R}^3$ case both the charges and masses are now determined from the leading asymptotic term in 
$\hat{\Phi}$.  Consequently, a given pattern of symmetry breaking can only be achieved by a particular choice 
of fundamental charges.  In fig.~\ref{fig01}, magnetic weights associated with minimal symmetry breaking are 
those lying on the lines $k_1=2k_2$ and $k_2=2k_1$.  This configuration is considered further in section 
\ref{su3pdcmon}.
\par  As pointed out in \cite{CK01}, the total energy is logarithmically divergent, such that the Bogomolny 
bound is
\begin{equation}
E\,=\,\frac{1}{2}\int_{\mathbb{R}^2\times\hat{S}^1}\text{tr}(\ast\hat{D}\hat{\Phi}\wedge\hat{D}\hat{\Phi})\,=\,\frac{1}{2}\int_{\rho=R}\text{tr}(\hat{\Phi}\ast\hat{D}\hat{\Phi})\,=\,\frac{\pi}{\beta}\sum_{i=1}^N\ell_i\left(\ell_i\log(R)+v_i\right)\label{energy}
\end{equation}
and we understand the Bogomolny equations to give a solution which minimises the energy in a region with $R$ 
large but finite.
\par As is done for the periodic instanton \cite{LL98}, it is useful to consider the holonomy in the periodic 
direction.  Explicitly, we are to solve the matrix equation
\begin{equation}
\partial_zV(\zeta,z)\,=\,\hat{\phi}\,V(\zeta,z)\label{holdef}
\end{equation}
with boundary condition $V(\zeta,0)=\bm{1}_N$, for $V(\zeta,\beta)$.  Under a gauge transformation with 
$\hat{g}=\hat{g}(\zeta,z)\in\text{SU}(N)$, the fields and holonomy transform as
\begin{IEEEeqnarray*}{rcl}
\hat{\Phi}\,&\mapsto&\,\hat{g}^{-1}\hat{\Phi}\hat{g}\nonumber\\
\hat{A}\,&\mapsto&\,\hat{g}^{-1}\hat{A}\hat{g}+\hat{g}^{-1}\text{d}\hat{g}\nonumber\\
V(\zeta,z)\,&\mapsto&\,\hat{g}^{-1}(\zeta,z)V(\zeta,z)\hat{g}(\zeta,0),\nonumber
\end{IEEEeqnarray*}
where $\hat{g}(\zeta,0)$ is introduced to ensure the boundary condition on $V(\zeta,z)$ is satisfied.  As 
long as $\hat{g}$ is strictly periodic, $\hat{g}(\zeta,\beta)=\hat{g}(\zeta,0)$, then the characteristic 
polynomial of $V(\zeta,\beta)$ is gauge invariant.  Asymptotically, using \eqref{asympfields}, the holonomy 
takes the form
\begin{equation}
V(\zeta,\beta)\,=\,\text{diag}\left(\zeta^{\ell_1}\text{e}^{\mathfrak{v}_1}(1+\mu_1\zeta^{-1}+\mathcal{O}(\rho^{-2})),\ldots\right).\label{asympholo}
\end{equation}
The analysis of the Bogomolny equations provided by \cite{CK01} establishes that the holonomy is in fact 
holomorphic, and is thus a polynomial in $\zeta$.
\subsection{Nahm Transform}\label{nahm}
It is shown in \cite{BvB89, CG84} that the Nahm transform provides a bijection between self-dual Yang-Mills 
fields on the torus $\hat{T}^4$ and the reciprocal torus $T^4$.  It is believed \cite{Jar04} that other 
self-dual Yang-Mills systems can be obtained by suitable rescalings of the tori.  In the present case, it is 
therefore expected that the Nahm dual to the monopole on $\mathbb{R}^2\times\hat{S}^1$ is a Hitchin system 
\cite{Hit87} on the `Hitchin cylinder' $\mathbb{R}\times S^1$ of period $2\pi/\beta$, and this was shown by 
\cite{CK01} to be the case.\footnote{The fact Hitchin equations are conformally invariant allows us to map 
solutions to other manifolds, including $\mathbb{R}^2$ or $S^2$.  We choose the cylinder to keep explicit the 
link with the Nahm transform.  This gains particular relevance when we make the comparison with doubly 
periodic instantons in section \protect\ref{singularities}.}  Following the notation of \cite{HW09,War05} the 
cylinder is parametrised by the coordinates $r\in\mathbb{R}$ and $t\in\mathbb{R}/(2\pi/\beta)\mathbb{Z}$, which 
are combined into a complex coordinate $s=r+\text{i}t$.  The Hitchin fields are of matrix rank $K$ and 
satisfy
\begin{equation}
F_{s\bar{s}}\,=\,-\tfrac{1}{4}[\Phi,\Phi^\dag]\qquad\qquad D_{\bar{s}}\Phi\,=\,\partial_{\bar{s}}\Phi+[A_{\bar{s}},\Phi]\,=\,0\label{hitchin}
\end{equation}
with $^\dag$ denoting Hermitian conjugation.  The monopole fields are recovered, up to a gauge, by finding 
solutions of the inverse Nahm equation,
\begin{equation}
\Delta\Psi\,=\,\begin{pmatrix}\bm{1}_K\otimes(2\partial_{\bar{s}}-z)+2A_{\bar{s}}&\bm{1}_K\otimes\zeta-\Phi\\\bm{1}_K\otimes\bar{\zeta}-\Phi^\dag&\bm{1}_K\otimes(2\partial_s+z)+2A_s\end{pmatrix}\,\Psi\,=\,0\label{invnahm}
\end{equation}
where $\Psi$ is a $(2K\times N)$ matrix subject to the normalisation condition
\begin{equation}
\int_{-\infty}^\infty\text{d}r\int_{-\pi/\beta}^{\pi/\beta}\text{d}t\,(\Psi^\dag\Psi)\,=\,\bm{N}_2.\label{normalisation}
\end{equation}
One can then, in principle, construct the monopole fields using
\begin{equation*}
\hat{\Phi}\,=\,\text{i}\int_{-\infty}^\infty\text{d}r\int_{-\pi/\beta}^{\pi/\beta}\text{d}t\,(r\Psi^\dag\Psi)\qquad\qquad\hat{A}_i\,=\,\int_{-\infty}^\infty\text{d}r\int_{-\pi/\beta}^{\pi/\beta}\text{d}t\,(\Psi^\dag\partial_i\Psi).
\end{equation*}
Gauge transformations $\hat{g}$ acting on the $\text{SU}(N)$ monopole fields and $g$ on the Nahm fields transform $\Psi$ as
\begin{equation}
\Psi(s;\zeta,z)\,\mapsto\,U(s)^{-1}\Psi(s;\zeta,z)\,\hat{g}(\zeta,z).\label{psigauge}
\end{equation}
where $U(s)=h\otimes g(s)$, with $h$ a constant $2\times2$ matrix serving to permute the entries in $\Delta$ 
and those of $\Psi$.  This freedom to rearrange makes it evident that it is irrelevant whether the 
derivatives $\partial_r$ and $\partial_t$ are introduced in the same or different entries of $\Delta$, the 
two configurations differing only by a choice of gauge.
\par Finally, it should be noted that in the $\beta\to0$ limit the Nahm transform is expected to be 
self-reciprocal, mapping between two Hitchin systems of different rank and boundary conditions.
\subsection{Spectral Data}\label{specdat}
The key observation of \cite{CK01,CK03} is that the characteristic equation of the $z$-holonomy, 
$\text{det}(w-V)=0$, relates monopole data to Nahm data through the parameter $w=\text{e}^{\beta s}$.  This 
provides a holomorphic curve {\bf S} in $\mathbb{C}\times\mathbb{C}^\ast$ known as the monopole spectral curve, 
which for an SU($N$) periodic monopole of charge $\bm{k}$ is
\begin{equation}
w^N+P_{1,k_1}(\zeta)w^{N-1}+\ldots+P_{N-1,k_{N-1}}(\zeta)w+(-1)^N\,=\,0\label{msc}
\end{equation}
where the $P_{i,k_i}(\zeta)$ denote polynomials in $\zeta$ with leading term proportional to $\zeta^{k_i}$.  
This relation shows that by performing a coordinate redefinition $w\mapsto w^{-1}$ the largest of the $k_i$ 
(if it is unique) can be chosen to lie in the first half of the entries in $\bm{k}$.  Referring to the 
SU($3$) case (fig.~\ref{fig01}), this amounts to identifying the regions on either side of the line 
$k_1=k_2$, and we will choose to work with the configurations below that line.
\par In addition to the monopole spectral curve \eqref{msc}, Cherkis \& Kapustin \cite{CK01,CK03} introduce a 
second, equivalent, spectral curve relating the coordinate on $\mathbb{R}^2$ in the monopole space to the 
characteristic equation of the Hitchin Higgs field $\Phi$,
\begin{equation}
\zeta^K-\text{tr}(\Phi)\zeta^{K-1}+\ldots+(-1)^K\text{det}(\Phi)\,=\,0,\label{hsc}
\end{equation}
where the intermediate terms are given by symmetric polynomials in the eigenvalues of $\Phi$.  By rewriting 
\eqref{msc} as a polynomial in $\zeta$, a comparison can be made with the coefficients of \eqref{hsc} to 
obtain $\Phi$.  In particular, it should be noted that $\text{det}(\Phi)$ will have singularities at finite 
$|r|$ if $K$ appears more than once in $\bm{k}$.  Smooth behaviour at large $|r|$ requires the introduction 
of singularities, both to the monopole and Hitchin fields.
\subsection{String Theory Setting}\label{strings}
The relation of periodic monopoles to compactified supersymmetric gauge theories is explained in detail in 
\cite{CK01,CK03,Kap98} and provides a physical context for the root structure presented in section 
\ref{monopoledata}.  The type IIB setup of interest consists of $N$ parallel D5-branes extended along the 
$x^0$-$x^5$ directions and $(N-1)$ stacks of $k_i$ D3-branes extended along the $x^0$-$x^2$ and $x^6$ 
directions ending on each of the $i^{\text{th}}$ pair of adjacent D5-branes, where $x^3$ is compactified on a 
circle.  From the point of view of the D5-brane system, each of the D3-branes is seen as a fundamental SU(2) 
periodic monopole of type $i$ localised in the $x^3$-$x^5$ directions of the D5-brane worldvolume, and 
translationally invariant along $x^0$-$x^2$.  Performing a $T$-duality in the $x^3$ direction returns a IIA system of D4-branes 
extended along $x^0$-$x^3$, $x^6$, ending on $N$ other D4-branes extended along $x^0$-$x^2$, $x^4$, $x^5$.  
The field equations on the ($x^3$, $x^6$)-cylinder are nothing other than the Hitchin equations of section 
\ref{nahm}.  The tension between the D4-branes causes them to deform, such that the $x^6$ direction of the 
cylinder becomes of infinite extent.
\begin{center}
\begin{tabular}{c|c c c c c c c c c c}
\phantom{NS5} & 0 & 1 & 2 & \!\large{\textcircled{\small{3}}}\! & 4 & 5 & 6 & 7 & 8 & 9 \\\hline
D5 & x & x & x & x & x & x & & & & \\
D3 & x & x & x & & & & x & & &
\end{tabular}
\end{center}
\begin{center}
$T_3\left\downarrow\rule{0cm}{0.6cm}\right.\phantom{T_3}$
\end{center}
\begin{center}
\begin{tabular}{c|c c c c c c c c c c}
\phantom{NS5} & 0 & 1 & 2 & \!\large{\textcircled{\small{3}}}\! & 4 & 5 & 6 & 7 & 8 & 9 \\\hline
D4 & x & x & x & & x & x & & & & \\
D4 & x & x & x & x & & & x & & &
\end{tabular}
\end{center}
\vspace{0.5cm}
\noindent Introducing $n_+$ and $n_-$ semi-infinite D3-branes ending on the first and $N^{\text{th}}$ D5-branes is 
equivalent to the introduction of Dirac singularities to the monopole system.  Compactifying the $x^6$ 
direction, such that the left and right D3-branes coincide, is equivalent to adding an $N^{\text{th}}$ root 
to the Lie algebra $\mathfrak{su}(N)$.  The series of dualities described above then leads to Hitchin 
equations on the 2-torus ($x^3$, $x^6$).  Such a system of singular monopoles and the relation of the torus 
to the Nahm data of the doubly periodic instanton will be discussed in section \ref{singularities}.
\section{Introducing the Spectral Approximation}\label{spap}
Due to the difficulty of finding exact solutions to the inverse Nahm operator \eqref{invnahm} and motivated 
by Ward's approximate $\bm{k}=(1)$ solution \cite{War05}, we will consider a construction based on the 
spectral curves (\ref{msc}, \ref{hsc}).  The following pargraphs describe the procedure to be followed and in 
the remainder of this section we use the results of \cite{War05} to illustrate the application and 
{\it r\'egime} of validity of the approximation.
\par Given an SU($N$) monopole with charge vector $\bm{k}$ it is straightforward to write down the spectral 
curves \eqref{msc} and \eqref{hsc}, where the polynomials $P_{i,k_i}(\zeta)$ can be expressed in terms of the 
data $\bm{\mathfrak{v}}$, $\bm{\mu}$.  We will be interested in the `spectral points', those values of 
$\zeta$ at which two or more of the eigenvalues of $V(\zeta,\beta)$ coincide.  These points are located by 
finding the zeroes of the discriminant $\mathcal{D}_{\bm{k}}$ of the polynomial in $w$ (as a function of 
$\zeta$).  For given $N$, the discriminant is obtained as the determinant of the rank $(2N-1)$ Sylvester 
matrix.  Our interest in the spectral points stems from the finding in the $\bm{k}=(1)$ case, discussed in 
section \ref{spcu}, that peaks in energy density are always located at the spectral points (though there 
appears to be no energy peak associated to two coincident spectral points, as will be seen for $\bm{k}=(2)$ 
in section \ref{charge2symmetries}).  It can be checked by explicit calculation for small $N$ that the 
highest power of $\zeta$ in $\mathcal{D}_{\bm{k}}$ is $2\sum_{i=1}^{N-1}k_i$, and we expect there to be this 
many spectral points.  We will see from various examples that away from the central region of the moduli 
space, the spectral points occur in pairs, forming $\sum_{i=1}^{N-1}k_i$ fundamental monopoles.
\par The spectral curve \eqref{msc} of the SU($N$) charge $\bm{k}$ periodic monopole has 
$2\sum_{i=1}^{N-1}(k_i+1)$ real coefficients.  It is expected \cite{CK03} that the complex coefficient of 
$\zeta^{k_i}$ in each of the polynomials $P_{i,k_i}(\zeta)$ is a parameter determined by the boundary data 
$\bm{\mathfrak{v}}$.  The centre of mass of the spectral points is factored out by choosing $\bm{\mu}$ such 
that the term of order $\zeta^{2\sum k_i-1}$ in $\mathcal{D}_{\bm{k}}$ vanishes, and we will say that such a 
monpole is centered.\footnote{It should be noted \protect\cite{CK02} that the infinite mass of a periodic 
monopole precludes variation of the centre of mass coordinates, and thus that one cannot define an uncentered 
moduli space.}  Overall, this yields $2\sum_{i=1}^{N-1}k_i-2$ real relative moduli, precisely half the number 
expected were we to consider the full three dimensional picture.  This suggests our approach is insensitive 
to relative $z$ and phase differences between the fundamental monopoles, such that its validity is expected 
to improve as the ratio of the monopole size to its period becomes large.  We will refer to the moduli 
appearing in the spectral curve as `reduced moduli', and will see in section \ref{ch2} that in the SU(2) 
charge $\bm{k}=(2)$ case they provide a geodesic submanifold of the full moduli space.
\subsection{SU(2) Charge 1 - Spectral Curve}\label{spcu}
We illustrate the procedure by reviewing the approximate construction of \cite{War05} for $\bm{k}=(1)$.  The 
spectral curves in this case are (recall that $\Phi$ is a matrix of rank $1$)
\begin{equation}
w^2-2(\zeta-a)w/C+1\,=\,0\qquad\qquad\zeta-\Phi\,=\,0.\label{speccharge1}
\end{equation}
The boundary data translates to $C=2\text{e}^{-\mathfrak{v}}$, $a=-\mu$, such that the Hitchin Higgs field is
\begin{equation*}
\Phi\,=\,a+C\cosh(\beta s)
\end{equation*}
while the Hitchin gauge potential $A_r$ can be set to zero by a gauge transformation and the Hitchin 
equations \eqref{hitchin} are satisfied trivially.  The inverse Nahm transform \eqref{invnahm} requires a 
solution of
\begin{equation}
\begin{pmatrix}2\partial_{\bar{s}}-z&\zeta-\Phi\\\bar{\zeta}-\Phi^\dag&2\partial_s+z\end{pmatrix}\begin{pmatrix}\Psi_+\\\Psi_-\end{pmatrix}\,=\,0\label{invnahmsu21}
\end{equation}
(such that $A_t$ is absorbed into $z$ and $a$ into $\zeta$).  For 
$(\zeta,\text{e}^{\beta s_0})\in\text{\bf{S}}$, $(\zeta-\Phi)$ will vanish at
\begin{equation}
\beta s\,=\,\pm\beta s_0\,=\,\pm\cosh^{-1}\left((\zeta-a)/C\right),\label{s0}
\end{equation}
such that away from the spectral curve,
\begin{equation*}
\zeta-\Phi\,=\,\pm\beta C(s\pm s_0)\sinh(\beta s_0)+\mathcal{O}(s\pm s_o)^2\,=\,\pm\beta(s\pm s_0)\xi+\mathcal{O}(s\pm s_o)^2
\end{equation*}
where $\xi^2=\zeta^2-C^2$.  As suggested by \cite{War05}, solutions to \eqref{invnahmsu21} are supported near 
the points $s=\pm s_0=\pm(r_0+\text{i}t_0)$ on the Hitchin cylinder.  The independent solutions take the form 
of Gaussian peaks localised at each of $\pm s_0$, assembled into
\begin{equation*}
\Psi\,\approx\,\mathcal{N}\begin{pmatrix}\xi E_-&|\xi|E_+\\-|\xi|E_-&\bar{\xi}E_+\end{pmatrix}
\end{equation*}
where
\begin{equation*}
\log(E_\pm(s))\,=\,-\tfrac{1}{2}\beta|\xi|\left((r\pm r_0)^2+(t\pm t_0)^2\right)-\text{i}zt
\end{equation*}
and we have chosen a different gauge to \cite{War05}, such that the monopole fields are traceless and 
explicitly independent of $z$.  Such a solution is valid when the peaks on $\mathbb{R}\times S^1$ are well 
separated, and narrow compared to the period of the cylinder.\footnote{If this were not the case we would not 
expect to find two independent solutions of \protect\eqref{invnahmsu21}, and there would be a dependence on 
the periodic coordinate $z$.  It is not possible to extract a factor of $\text{e}^{\pm\text{i}zt}$ from 
solutions which are not narrow relative to the circumference of the cylinder while simultaneously preserving 
the periodicity condition.}  These conditions are ensured if we stay away from the spectral points 
$\zeta=\pm C$,
\begin{equation}
|\zeta^2-C^2|\,\gg\,\frac{\beta^2}{16\pi^2}.\label{spectralregion}
\end{equation}
In this region, the normalisation factor $\mathcal{N}$ is determined from \eqref{normalisation} to be 
$|\mathcal{N}|^2=\beta/(2\pi|\xi|)$ and the monopole fields, after a gauge transformation 
$\hat{g}=\text{exp}(\frac{1}{4}\log(\bar{\xi}/{\xi})\sigma_3)$ are
\begin{equation}
\hat{\Phi}\,=\,\text{i}r_0\sigma_3\qquad\qquad\hat{A}_{z}\,=\,-\text{i}t_0\sigma_3\label{fields}
\end{equation}
\begin{equation*}
\hat{A}_\zeta\,=\,\frac{\zeta}{4\xi^2}\,\text{e}^{-\beta|\xi||s_0|^2}\sigma_1\qquad\qquad\hat{A}_{\bar{\zeta}}\,=\,-\hat{A}_\zeta^\dag
\end{equation*}
where to ensure we remain on the correct branch we choose
\begin{equation*}
|s_0|^2\,=\,\inf_{n\in\mathbb{Z}}\left(r_0^2+(t_0+n/2)^2\right).
\end{equation*}
It is important to note that the fact the monopole Higgs field can be read off directly from the spectral 
curve \eqref{speccharge1} via $s_0$ \eqref{s0} is not simply a restatement of the boundary conditions, as use 
has also been made of the fact the coefficients in $w$ of the spectral curve are polynomials in $\zeta$, 
which encode the moduli in a particular way \cite{CK03}.  This result will be used in sections 
\ref{ch2spap}, \ref{su3pdcmon} and \ref{singularities} when we discuss the charge $2$, SU(3) and singular 
periodic monopoles.
\par It is useful to combine the fields \eqref{fields} into $\text{i}\hat{\phi}=\hat{\Phi}-\text{i}\hat{A}_z$ 
and $\hat{a}=\hat{A}_\zeta\text{d}\zeta+\hat{A}_{\bar{\zeta}}\text{d}\bar{\zeta}$ (see \eqref{asympfields}).  
We note that $\hat{a}$ approaches zero exponentially away from the spectral points $\zeta=\pm C$, and the 
fields are Abelian and trivially satisfy Hitchin equations in this limit, suggesting that they are truly 
two dimensional.  Noting that $|s_0|$ has dimensions of $\beta^{-1}$, we conjecture that in the limit 
$\beta\to0$ a solution is provided by
\begin{equation*}
\hat{\phi}\,=\,s_0\sigma_3\qquad\qquad\hat{a}\,=\,0,
\end{equation*}
which satisfies the Bogomolny equations with the correct boundary conditions \eqref{asympfields}.  As will be 
seen in section \ref{secmetric}, this approximation also leads to the correct asymptotic behaviour of the 
moduli space.
\subsection{Charge 1 - Energy}\label{ch1en}
\begin{figure}
\centering
\includegraphics[width=0.8\textwidth]{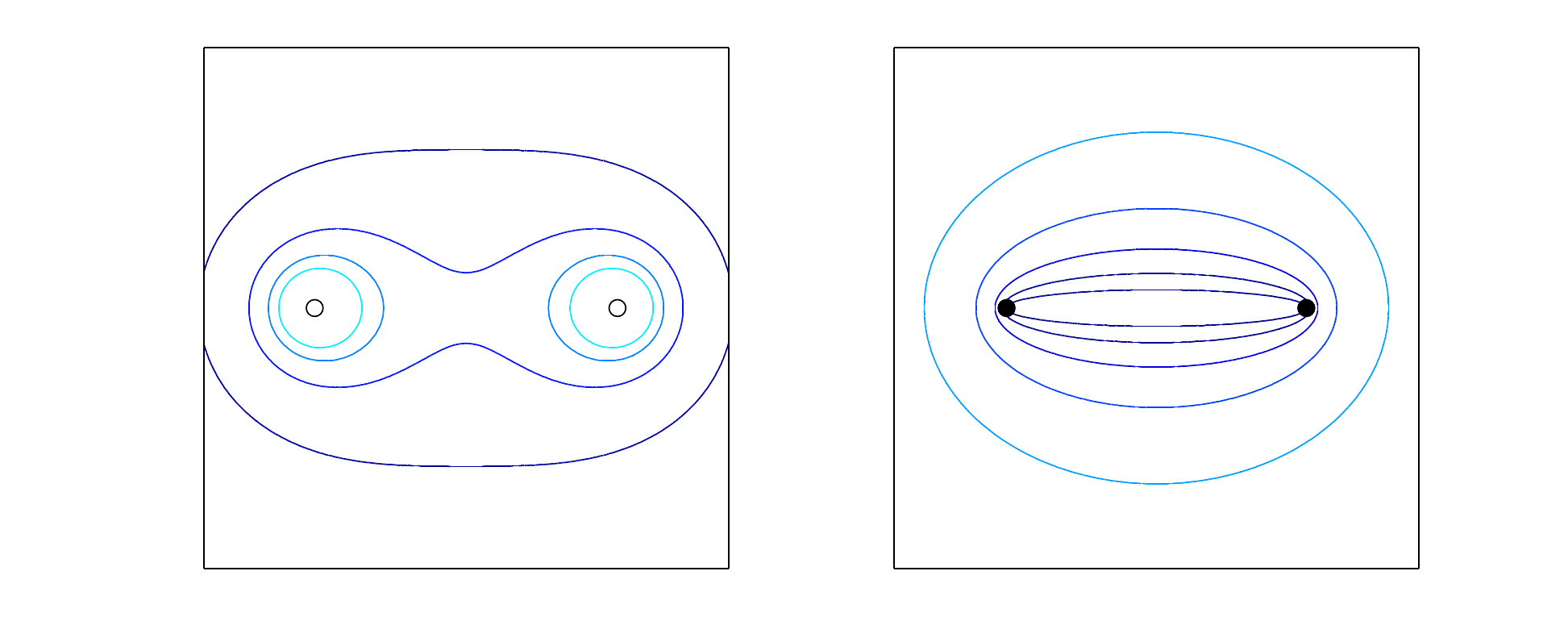}
\caption{An SU(2) monopole.  On the left is a contour plot of the energy density \protect\eqref{endens} and 
on the right $\log(\text{disc}(\hat{\Phi})+0.001)$, where the discriminant is defined as the squared 
difference of the eigenvalues of $\hat{\Phi}$.  It is zero on a line joining the spectral points, whose 
locations are indicated by black dots on the right hand diagram.  Note the loss of axial 
symmetry.}\label{fig02}
\end{figure}

It is convenient to rewrite the energy density, the integrand of \eqref{energy} in terms of just the Higgs 
field by using the Bianchi identity \cite{MS07,War81},
\begin{equation}
\mathcal{E}\,=\,\tfrac{1}{4}\,\nabla^2\,|\text{tr}(\hat{\Phi}^2)|.\label{nablaenergy}
\end{equation}
The Higgs field \eqref{fields} is
\begin{equation}
\hat{\Phi}\,=\,\frac{\text{i}}{\beta}\,\Re\left(\cosh^{-1}\left(\frac{\zeta}{C}\right)\right)\sigma_3\,=\,\frac{\text{i}}{\beta}\,\log\left|\frac{\zeta}{C}+\sqrt{\left(\frac{\zeta}{C}\right)^2-1}\right|\sigma_3,\label{higgssu21}
\end{equation}
giving an energy density
\begin{equation}
\mathcal{E}_1\,=\,\frac{1}{\beta^2|\xi|^2}\,=\,\frac{1}{\beta^2}\frac{1}{\sqrt{\rho^4-2\rho^2C^2\cos(2\theta)+C^4}}\label{endens}
\end{equation}
whose contours trace out Cassini ovals (fig.~\ref{fig02}) and is peaked at the spectral points, whose 
separation by $2C$ allows us to interpret $C$ as the characteristic size of the monopole.
\par We next use the divergence theorem to compute the total energy in a region with $\rho=R\gg C$
\begin{equation*}
V_1\,=\,\tfrac{1}{4}\int\!\!\!\!\int\!\!\!\!\int_{\rho\leq R}\nabla^2|\text{tr}(\hat{\Phi})^2|\rho\text{d}\rho\text{d}\theta\text{d}z\,=\,\tfrac{1}{4}R\beta\int_{\rho=R}\left(\partial_\rho|\text{tr}(\hat{\Phi})^2|\right)\text{d}\theta
\end{equation*}
and note that the leading term of the integrand at large $\rho$ is
\begin{equation*}
\partial_\rho|\text{tr}(\hat{\Phi})|^2\,\sim\,\frac{4}{\rho\beta^2}\log\left(\frac{2\rho}{C}\right),
\end{equation*}
resulting in
\begin{equation*}
V_1\,=\,\int\!\!\!\!\int\!\!\!\!\int_{\rho\leq R}\mathcal{E}_1\rho\text{d}\rho\text{d}\theta\text{d}z\,=\,\frac{2\pi}{\beta}\log\left(\frac{2R}{C}\right)
\end{equation*}
in agreement with \eqref{energy}.  Finally, we note that the Higgs field \eqref{higgssu21} vanishes along a 
line between the spectral points, fig.~\ref{fig02}, as can be seen in the numerical study of \cite{DK05}.  We 
will see in section \ref{singularities} that this observation survives for higher charges.
\section{Charge 2 - Spectral Approximation}\label{ch2spap}
In this section we apply the spectral approximation to the SU(2) monopole of charge $\bm{k}=(2)$, which has 
two real reduced moduli.  Using symmetries of the spectral curves this can be reduced to two one-parameter 
families, though we withhold showing that this two dimensional reduced moduli space is itself a geodesic 
submanifold of the full four dimensional moduli space until section \ref{fullsymms}.
\par In the limit in which the approximation becomes exact it is possible to compute a metric on the two 
dimensional reduced moduli space.  Its asymptotic form agrees with the ALG metric of \cite{CK02}, allowing 
numerical integration of non-trivial geodesics, which will be considered both in the monopole space and on 
the dual cylinder.  Finally, we will introduce a new solution of the rank $2$ Hitchin system \cite{Har} with 
the same spectral limit as that of \cite{HW09}, and briefly compare their scattering properties.
\subsection{Spectral Approximation}\label{charge2specaprox}
The general form of the monopole spectral curve \eqref{msc} of the charge $\bm{k}=(2)$ periodic monopole is
\begin{equation}
w^2+P_{1,2}(\zeta)w+1\,=\,0\qquad\text{with}\qquad P_{1,2}(\zeta)\,=\,-\left(2\zeta^2-2BC\zeta-K\right)/C.\label{speccharge2}
\end{equation}
where $B,C\in\mathbb{C}$.  The spectral points are located at the values of $\zeta$ where $(P_{1,2}(\zeta))^2=4$.  Fixing the centre of 
mass at the origin, we expect energy peaks at the four points
\begin{equation*}
\zeta\,=\,\pm\sqrt{K/2\pm C}
\end{equation*}
(where the $\pm$ signs are independent).\footnote{Note that to regain the $\bm{k}=(1)$ limit we should 
instead fix $B$ and $K$ and set $|C|\to\infty$.}  As in the $\bm{k}=(1)$ case, $C$ is a parameter fixed by 
the boundary conditions, while $K$ is a complex modulus.  For $|K|\gg2|C|$ the spectral points occur in two 
pairs which are interpreted as fundamental monopoles of size $|C\sqrt{2/K}|$ separated by a distance 
$|\sqrt{2K}|$.  It is noteworthy that the fundamental monopoles get smaller as they are separated, an effect 
of the long range Higgs field.
\par Motivated by \eqref{fields} we assume the monopole Higgs field is given by 
$\hat{\Phi}=\text{i}\Re(s_0)\sigma_3$, where $s_0$ is obtained by rearranging the spectral curve,
\begin{equation*}
\hat{\Phi}\,=\,\frac{\text{i}}{\beta}\,\Re\left(\cosh^{-1}\left(\frac{2\zeta^2-K}{2C}\right)\right)\sigma_3
\end{equation*}
and compute the energy in a region with $|\zeta|<R$ using \eqref{nablaenergy} to find
\begin{equation*}
V_2\,=\,\frac{4\pi}{\beta}\log\left(\frac{2R^2}{C}\right),
\end{equation*}
again in agreement with \eqref{energy}.  Applying the divergence theorem to $\partial_K\mathcal{E}$ for large 
$\rho$,
\begin{equation*}
\partial_K\mathcal{E}\,\propto\,\partial_K\partial_\rho|\text{tr}(\hat{\Phi}^2)|\,\sim\,\rho^{-3}\log(\rho),
\end{equation*}
shows that, as hoped, the total energy is independent of the modulus $K$.
\subsection{Aside - Symmetric Charge k}\label{chargekenergy}
The spectral curve of the $\mathbb{Z}_{2k}$-symmetric charge $\bm{k}=(k)$ monopole is
\begin{equation*}
C\cosh(\beta s)\,=\,\zeta^k\qquad\Rightarrow\qquad\hat{\Phi}\,=\,\frac{\text{i}}{\beta}\,\Re\left(\cosh^{-1}\left(\frac{\zeta^k}{C}\right)\right)\sigma_3,
\end{equation*}
from which the energy density \eqref{endens} is
\begin{equation*}
\mathcal{E}_k\,=\,\frac{k^2}{\beta^2}\frac{\rho^{2k-2}}{\sqrt{\rho^{4k}-2C^2\rho^{2k}\cos(2k\theta)+C^4}},
\end{equation*}
where we note that the energy density at the origin vanishes for all $k>1$.
\par
\begin{figure}
\centering
\includegraphics[width=8cm]{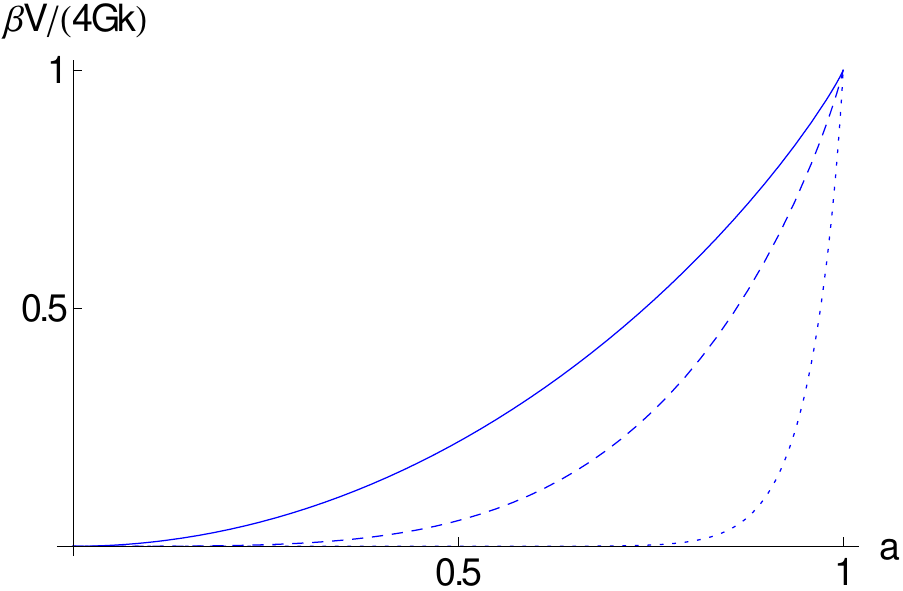}
\caption{Normalised energy per unit charge enclosed in a period cylinder of radius $aC^{1/k}$ for various 
values of the charge $k$.  Solid line: $k=1$, dashed: $k=2$, dotted: $k=10$.  The energy density is 
increasingly located on a shell of radius $\rho=C^{1/k}$.}\label{fig03}
\end{figure}

 The total energy is again in agreement with \eqref{energy}, while the energy per unit charge in the 
region $0\leq\rho\leq aC^{1/k}$ is (note that the spectral points are located on a circle of radius 
$\rho=C^{1/k}$)
\begin{IEEEeqnarray}{rcl}
\frac{V_k}{k}\,(0\leq\rho\leq aC^{1/k})\,&=&\,\frac{\pi a^{2k}}{\beta}\!\!\!\!\phantom{F}_3F_2\left(\tfrac{1}{2},\tfrac{1}{2},\tfrac{1}{2};1,\tfrac{3}{2};a^{4k}\right)\label{enpercharge}\\
&=&\,\left\{\begin{array}{lll}\pi a^{2k}\left(1+\mathcal{O}(a^{4k})\right)/\beta&&(a<1)\\4G/\beta&&(a=1)\end{array}\right.\nonumber
\end{IEEEeqnarray}
where $\!\!\!\!\phantom{F}_3F_2$ is the generalised hypergeometric function, $G\approx0.916$ is Catalan's 
constant and we have used the following identities for the elliptic integral $\bm{K}(\kappa)$ \cite{GR94}:
\begin{equation*}
\bm{K}(\kappa)\,=\,\int_0^{\pi/2}\frac{1}{\sqrt{1-2\kappa\cos(2\alpha)+\kappa^2}}\,\text{d}\alpha\qquad\qquad(\kappa<1),
\end{equation*}
\begin{equation}
4ab\int_0^z\kappa^{2ab-1}\bm{K}(\kappa^b)\,\text{d}\kappa\,=\,\pi z^{2ab}\!\!\!\!\phantom{F}_3F_2\left(\tfrac{1}{2},\tfrac{1}{2},a;1,a+1;z^{2b}\right).\label{hypergeom}
\end{equation}
Fig.~\ref{fig03} shows the total energy in a period cylinder, \eqref{enpercharge}, is increasingly located 
at its edge as $k$ is increased.  An expansion of the fields at small and large $\rho$ yields
\begin{equation*}
\left\{\begin{array}{lll}-\text{i}\beta\hat{\Phi}\,\sim\,(\rho^k/C)\sin(k\theta)\sigma_3&&(\rho^k\ll C),\\-\text{i}\beta\hat{\Phi}-\log\left(2\rho^k/C\right)\sigma_3\,\sim\,(2\rho^k/C)^{-2}\cos(2k\theta)\sigma_3&&(\rho^k\gg C).\end{array}\right.
\end{equation*}
These results resemble those found for spherical magnetic bags of large charge, as first studied by 
\cite{Bol06}, and it is interesting to see evidence of a `magnetic cylinder' with similar properties.
\subsection{Symmetries}\label{charge2symmetries}
Geodesic submanifolds of the two dimensional reduced moduli space are obtained by looking for symmetries in 
the spectral curve \eqref{speccharge2}.  Fixing the parameters $B=0$ and $C\in\mathbb{R}$, we impose 
invariance of \eqref{speccharge2} under a reflection symmetry in the line $\theta=\alpha/2$, encoded by the 
map $\zeta\mapsto\text{e}^{\text{i}\alpha}\bar{\zeta}$.  This requires that we simultaneously map 
$w\mapsto\text{e}^{-2\text{i}\alpha}\bar{w}$ ($t\mapsto-t-2\alpha/\beta$) and 
$K\mapsto\text{e}^{2\text{i}\alpha}\bar{K}$.  The original spectral curve \eqref{speccharge2} is recovered by 
complex conjugation as long as $\alpha$ is chosen to be $0$ or $\pi/4$.  These choices of $\alpha$ correspond 
to the one parameter families $K\in\mathbb{R}$ and $K\in\text{i}\mathbb{R}$, respectively.  In section 
\ref{ch2} it will be shown that the reduced moduli provide a geodesic submanifold of the full four 
dimensional moduli space, allowing us to consider the above one parameter families as geodesics.  The 
definition of a metric on the reduced moduli space will be considered in the following subsection.
\par
\begin{figure}
\begin{minipage}{0.485\textwidth}
\centering
\includegraphics[width=0.95\textwidth]{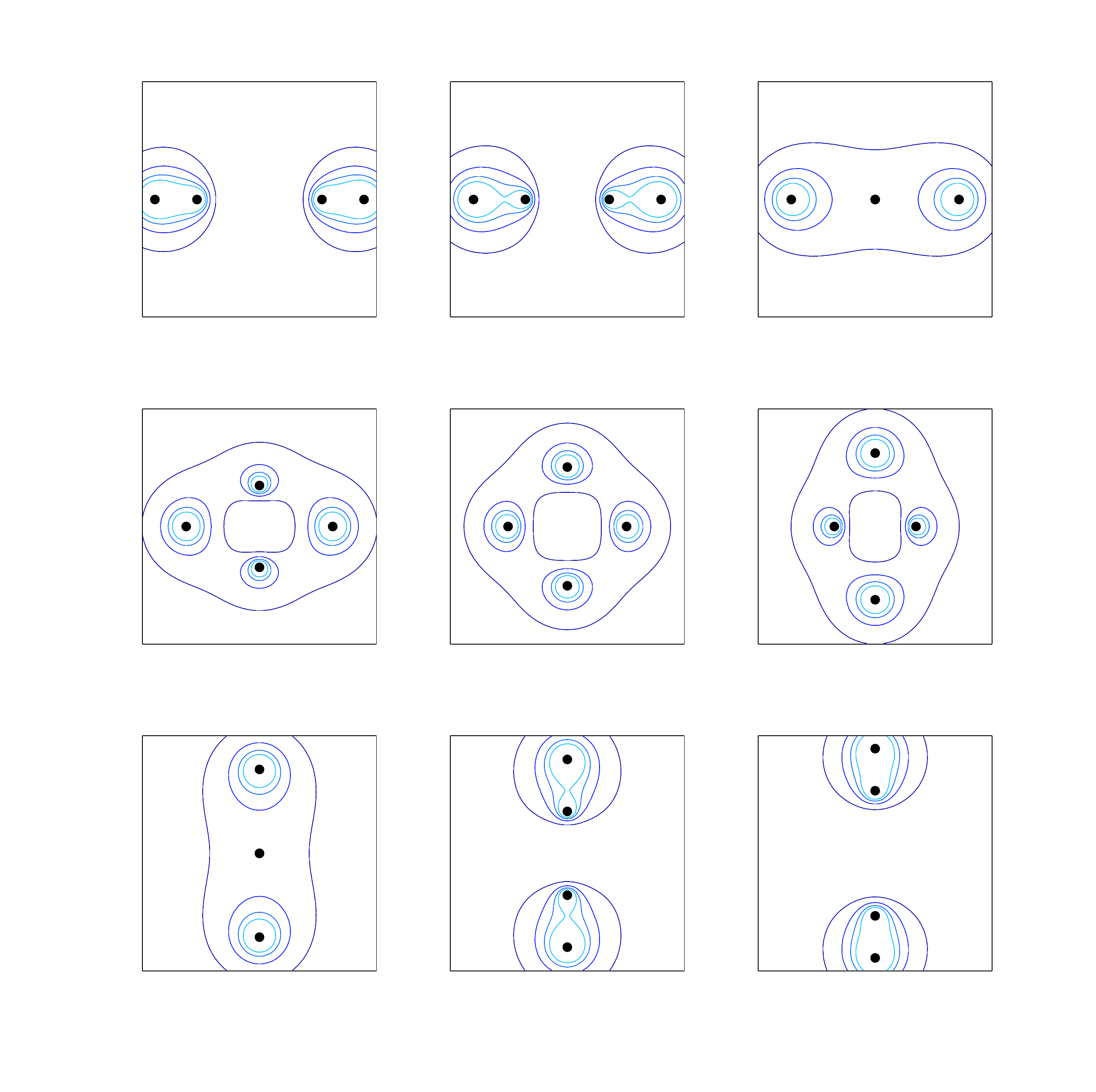}
\end{minipage}
\begin{minipage}{0.485\textwidth}
\centering
\includegraphics[width=0.95\textwidth]{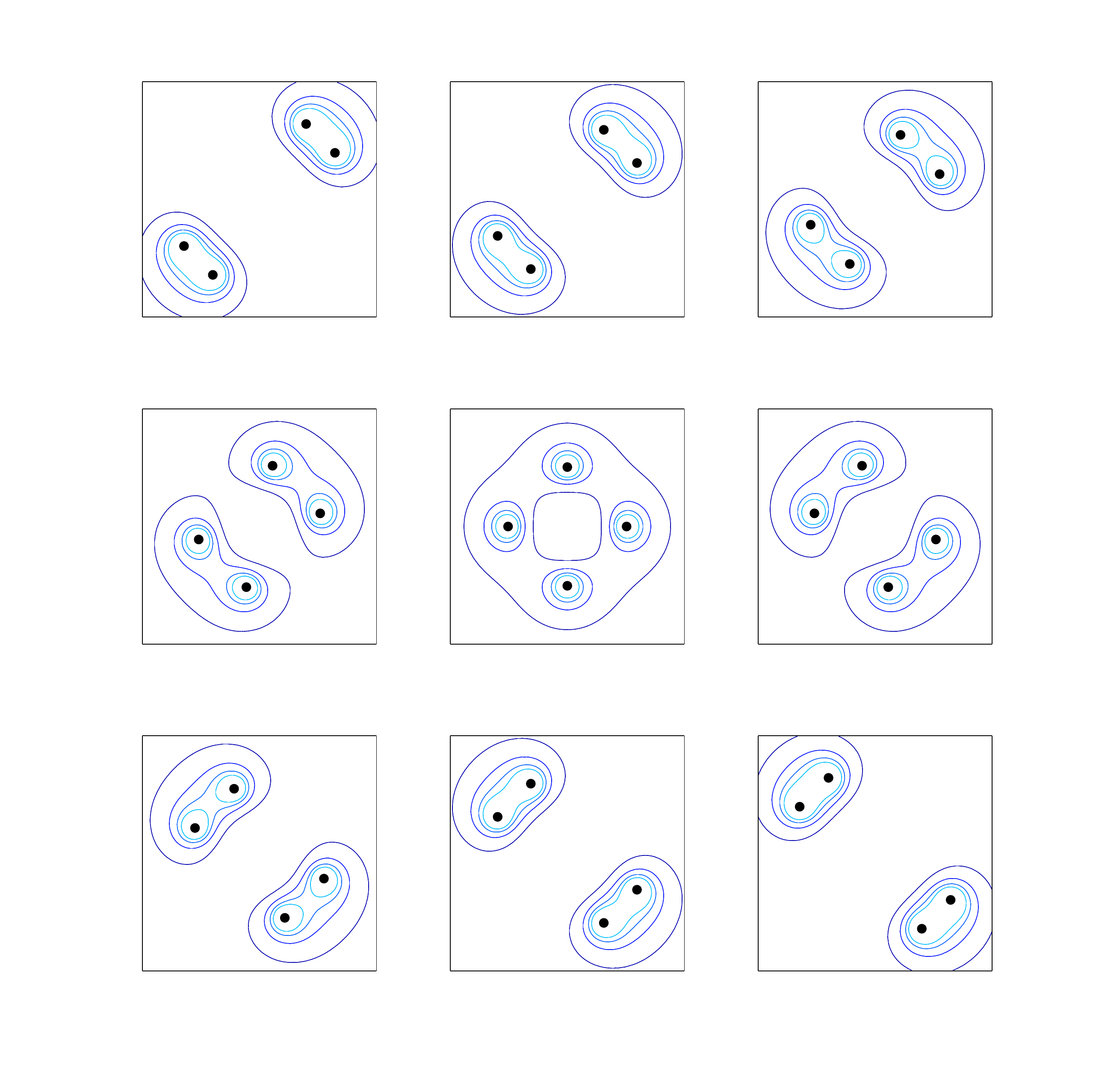}
\end{minipage}
\vspace{-0.5cm}
\caption{Energy density contour plots.  Left: geodesic with $K\in\mathbb{R}$ (to be read from left to right 
and top to bottom).  Right: geodesic with $K\in\text{i}\mathbb{R}$.  The central symmetric configurations 
have $K=0$, while those with just two energy peaks have $K=\pm2$.  It is noteworthy that the axial symmetry 
of the `doughnut' charge $2$ monopole in $\mathbb{R}^3$ is replaced by a discrete symmetry.  The spacing 
between snapshots is taken relative to the metric defined in section 
\protect\ref{secmetric}.}\label{fig04}
\end{figure}

 More information about these geodesics can be gleaned from considering the $\pi/2$ rotation symmetry 
$\zeta\mapsto\mathrm{i}\zeta$, which requires $w\mapsto-w$ ($t\mapsto t+\pi/\beta$) and $K\mapsto-K$.  For 
the one parameter families found above, passing through $K=0$ leads to the right-angled scattering processes 
shown in fig.~\ref{fig04}.  Particularly interesting points in the moduli space are $K=\pm2C$, where two 
of the spectral points coincide at the origin (although there is no energy peak associated with them) and 
$K=0$, where the dihedral $D_2$ symmetry is enhanced to $D_4$.  This is nothing but the symmetric 
configuration considered in section \ref{chargekenergy}.  For $K/C\in[-2,2]$ the fundamental monopoles lose 
their individual identities and the discriminant vanishes on a cross shape joining the four peaks.
\subsection{Metric}\label{secmetric}
\subsubsection{Definition}
We use the general formalism for obtaining the moduli space metric from the variation of the fields (see, for 
example, \cite{MS07}).  For $z$-independent fields the metric is given by
\begin{equation*}
g\,=\,\tfrac{1}{2}\dot{K}\dot{\bar{K}}\int_{\mathbb{R}^2}\text{tr}\left(\partial\hat{\phi}\bar{\partial}\hat{\phi}^\dag+\partial\hat{\phi}^\dag\bar{\partial}\hat{\phi}-4\partial\hat{a}_{\bar{\zeta}}\bar{\partial}\hat{a}_\zeta-4\partial\hat{a}_\zeta\bar{\partial}\hat{a}_{\bar{\zeta}}\right)\text{d}^2x
\end{equation*}
where it is understood that the fields satisfy the gauge condition
\begin{equation}
2\left(D_\zeta\partial(\hat{a}_{\bar{\zeta}})+D_{\bar{\zeta}}\partial(\hat{a}_\zeta)\right)\,=\,\tfrac{1}{2}[\hat{\phi},\partial(\hat{\phi}^\dag)]+\tfrac{1}{2}[\hat{\phi}^\dag,\partial\hat{\phi}]\label{ortho}
\end{equation}
which arises as a dimensional reduction of the equivalent condition for instantons, 
$D_\mu(\partial A_\mu)=0$.  Here $\partial$ indicates differentiation with respect to $K$, and $\dot{}$ is 
differentiation with respect to an affine time $\tau$.
\par From \eqref{fields} there is a centered charge $2$ solution of the Bogomolny equations with
\begin{equation*}
\beta\hat{\phi}\,=\,\cosh^{-1}\left(\frac{2\zeta^2-K}{2C}\right)\sigma_3\qquad\qquad\hat{a}\,=\,0,
\end{equation*}
valid sufficiently far from the spectral points, for which the orthogonality condition \eqref{ortho} holds 
trivially.  As discussed in section \ref{spcu}, it will be assumed that this becomes exact in the limit of 
$z$-independence.  It follows that the metric is given by
\begin{equation}
g\,=\,\frac{1}{4\beta^2}\dot{K}\dot{\bar{K}}\int\left((\zeta^2-K/2)^2-C^2\right)^{-1/2}\left((\bar{\zeta}^2-\bar{K}/2)^2-C^2\right)^{-1/2}\,\rho\text{d}\rho\text{d}\theta.\label{metricintegral}
\end{equation}
For given $K$ the integral can be written in terms of products of distances to the four spectral points, 
which are located at $\zeta_i(K)=\pm\sqrt{K/2\pm C}$, defining the conformal factor $\eta(K)$,
\begin{equation*}
g\,=\,\frac{1}{4\beta^2}\dot{K}\dot{\bar{K}}\int\frac{1}{|\zeta-\zeta_1||\zeta-\zeta_2||\zeta-\zeta_3||\zeta-\zeta_4|}\,\rho\text{d}\rho\text{d}\theta\,=\,\eta\dot{K}\dot{\bar{K}}.
\end{equation*}
We note that due to holomorphicity of the Higgs field $\hat{\phi}$ the moduli space is a 
one-complex-dimensional Hermitian manifold.  As expected for a complex submanifold of the 
four-real-dimensional hyper-K\"ahler moduli space \cite{CK02}, $g$ is indeed K\"ahler, with K\"ahler 
potential proportional to $\int|\text{tr}(\hat{\phi}^2)|\rho\text{d}\rho\text{d}\theta$.
\subsubsection{Asymptotics}
The integral in \eqref{metricintegral} can be computed in the limit in which the monopoles are well 
separated, $|K|\gg2|C|$.  Two of the peaks are placed near the origin, at $\zeta=\pm\epsilon$, and the others 
are centered at some large $R$ along the $x$-axis (for simplicity we consider 
$K=k\text{e}^{\text{i}\alpha}\in\mathbb{R}$).  Integrating out to some $\rho_0$ ($R\gg\rho_0\gg\epsilon$),
\begin{equation*}
\eta\,\sim\,\frac{1}{R^2}\int_0^{\rho_0}\frac{1}{|\zeta+\epsilon||\zeta-\epsilon|}\,\rho\text{d}\rho\text{d}\theta.
\end{equation*}
This integrand is identical to that of \eqref{endens}, so
\begin{equation*}
\eta\,\sim\,\frac{1}{R^2}\log(2\rho_0/\epsilon).
\end{equation*}
We recall from section \ref{charge2specaprox} that the separation and size of the fundamental monopoles in 
this limit are, respectively,
\begin{equation*}
R\,=\,\sqrt{2k}\qquad\qquad\epsilon\,=\,C(2k)^{-1/2}\,=\,C/R,
\end{equation*}
allowing us to express the metric either in terms of $k$ or the monopole separation $R$,
\begin{equation*}
g\,\sim\,\frac{1}{k}\left(\log(k)+c\right)\dot{k}^2\,\sim\,\left(\log(R)+c'\right)\dot{R}^2.\label{asympmec}
\end{equation*}
The latter agrees, up to prefactors, with the asymptotic metric computed in \cite{CK02}, which is an ALG 
metric of limiting Gibbons-Hawking type \cite{GH78}.  The constants $c$ and $c'$ depend on the upper limit of 
integration $\rho_0$ and are related to the redefinition of $\mathfrak{v}$ performed in \cite{CK02} when a chain 
of $n$ monopoles is studied in the limit of $n\to\infty$.
\subsubsection{Integration}
\begin{figure}
\centering
\includegraphics[width=10cm]{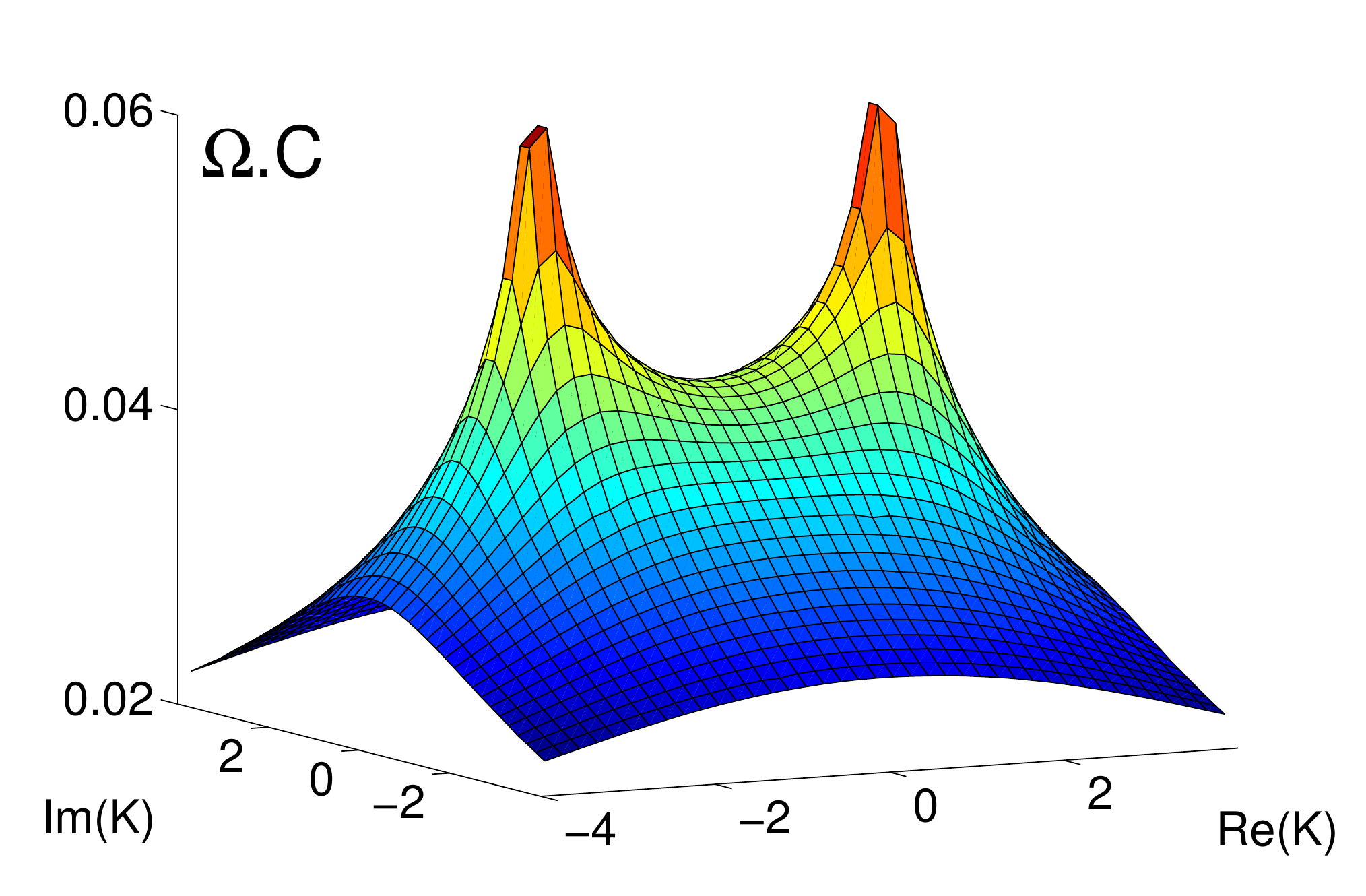}
\caption{Surface plot of the conformal factor for the relative reduced moduli space in the limit of large 
monopole size to period ratio.  Peaks are at $K=\pm2C$.}\label{fig05}
\end{figure}

There are three specific values of $K$ at which evaluation of the conformal factor $\eta$ can be performed 
analytically (see fig.~\ref{fig04} for the relevant monopole configurations),
\begin{IEEEeqnarray}{lll}
K\,=\,0&\qquad\qquad&\eta\,=\,\frac{1}{32\pi\beta^2C}\,\left(\Gamma\left(\tfrac{1}{4}\right)\right)^4\nonumber\\
K\,\to\,\pm2C&\qquad\qquad&\eta\,\sim\,-\frac{\pi}{8\beta^2C}\log\left(|K\mp2C|\right)\label{eta}
\end{IEEEeqnarray}
where, for $K=0$, use has been made of \eqref{hypergeom}.  The integral diverges at $K=\pm2C$, when two of 
the spectral points coincide and there is a double pole in the integrand.  We employ these results to ensure 
a correct numerical implementation of the integral for general $K$, and the result is shown in 
fig.~\ref{fig05}.  Further evidence for this metric will be provided in \cite{MW}.
\par Using polar coordinates $K=k\text{e}^{\text{i}\alpha}$ the geodesic equations are
\begin{IEEEeqnarray}{rcl}
2\eta k^2\ddot{\alpha}+(\partial_\alpha\eta)(k^2\dot{\alpha}^2-\dot{k}^2)+2(\partial_k\eta)k^2\dot{\alpha}\dot{k}+4\eta k\dot{\alpha}\dot{k}\,&=&\,0\nonumber\\
2\eta\ddot{k}+(\partial_k\eta)(\dot{k}^2-k^2\dot{\alpha}^2)+2(\partial_\alpha\eta)\dot{\alpha}\dot{k}-2\eta k\dot{\alpha}^2\,&=&\,0\label{geoeqs}
\end{IEEEeqnarray}
where $\dot{}$ denotes differentiation with respect to the parameter time $\tau$.  In particular, there are 
geodesics with $\dot{\alpha}=0$, for which the geodesic equations become $\partial_\alpha\eta=0$ and
\begin{equation}
2\eta\ddot{k}+(\partial_k\eta)\dot{k}^2\,=\,0\qquad\Rightarrow\qquad\int\sqrt{\eta}\,\text{d}k\,=\,b_1\tau+b_2,\label{geoeq}
\end{equation}
where $b_1$ and $b_2$ are constants of integration.  As can be seen from fig.~\ref{fig05} such geodesics 
are only possible for $\alpha=0,\pi/2$, which are precisely the geodesic submanifolds $K\in\mathbb{R}$ and 
$K\in\text{i}\mathbb{R}$ obtained by symmetry arguments in section \ref{charge2symmetries}.
\par The logarithmic behaviour of $\eta$ in the vicinity of $K=\pm2C$ \eqref{eta}, combined with the implicit 
expression for $k(\tau)$ \eqref{geoeq}, is sufficient to show that geodesics cross the points $K=\pm2C$ in 
finite parameter time.
\subsubsection{New Geodesics}
\begin{figure}
\begin{minipage}{0.485\linewidth}
\centering
\vspace{-0.3cm}
\includegraphics[width=0.75\linewidth]{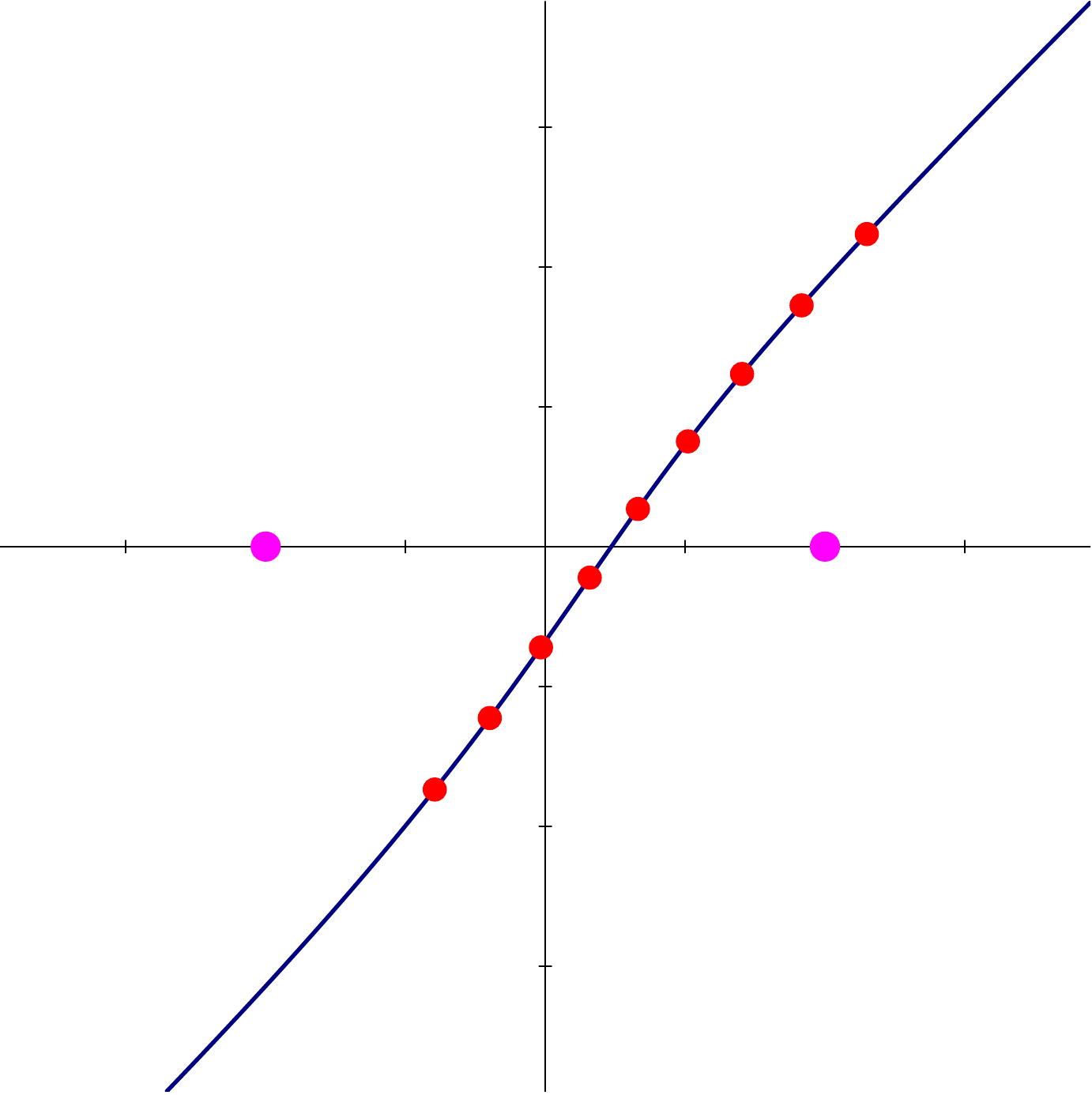}
\end{minipage}
\begin{minipage}{0.485\linewidth}
\centering
\includegraphics[width=0.95\linewidth]{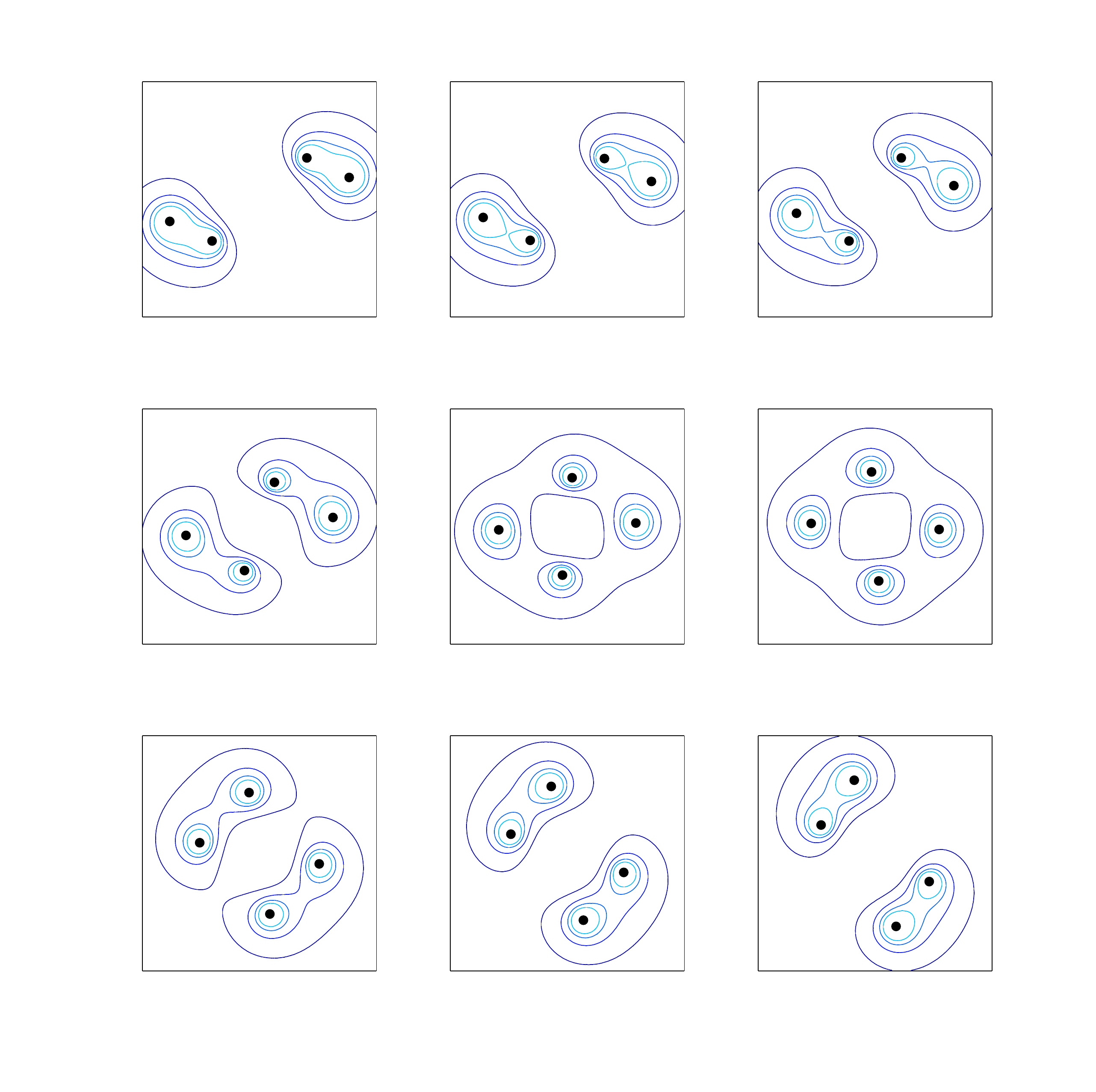}
\end{minipage}
\vspace{-0.5cm}
\caption{Geodesic for initial condition $K/C=5(1+\text{i})$, $\dot{K}/C=-0.03(1+\text{i})$ with step size 
$0.03$.  The left hand plot displays the geodesic on the $K$-plane (with shaded circles at $K/C=\pm2$).  Tick 
marks every $722$ timesteps indicate the positions of the energy density snapshots displayed to the 
right.}\label{fig06}
\end{figure}

\begin{figure}
\begin{minipage}{0.485\linewidth}
\centering
\vspace{-0.3cm}
\includegraphics[width=0.75\linewidth]{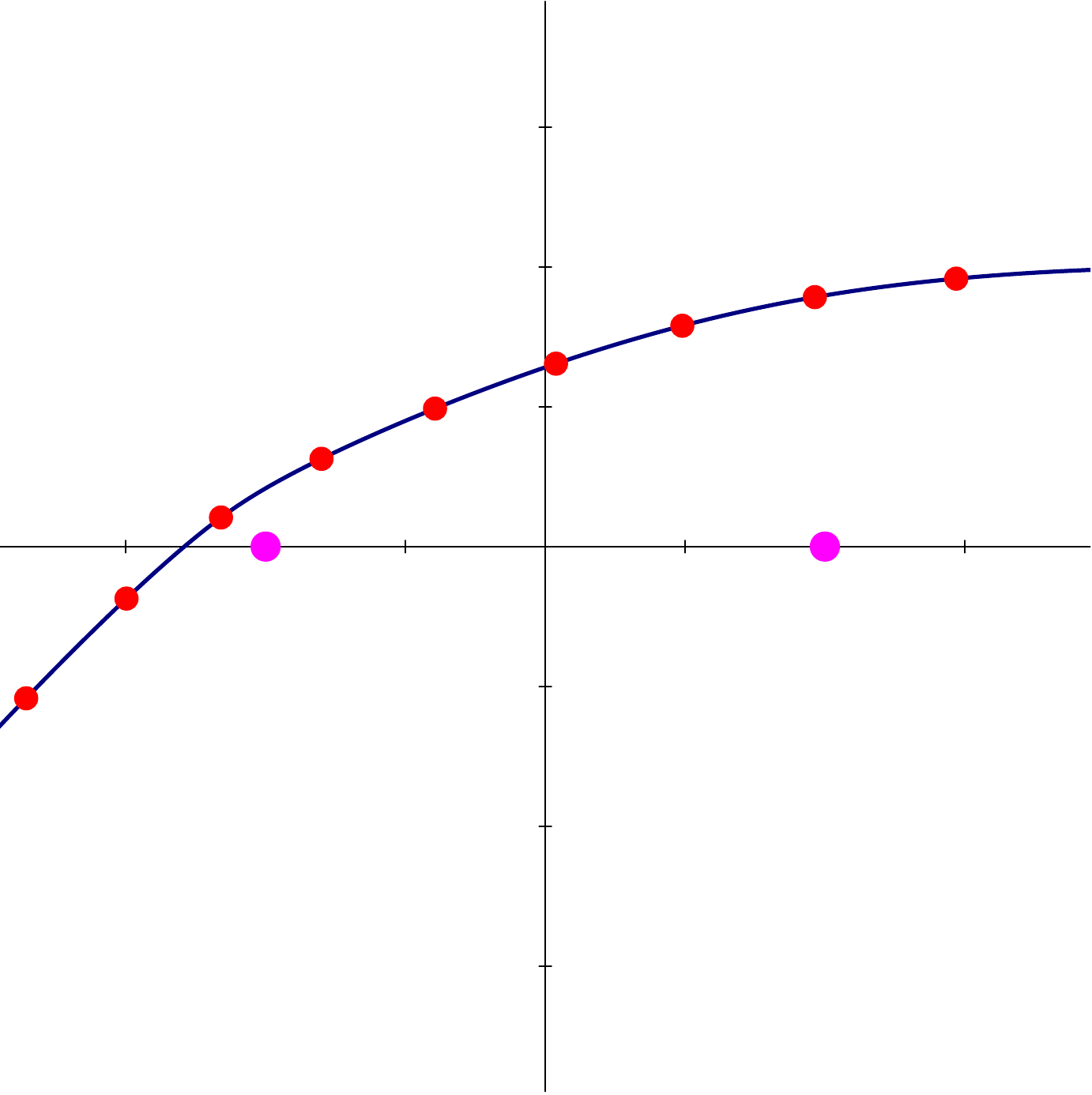}
\end{minipage}
\begin{minipage}{0.485\linewidth}
\centering
\includegraphics[width=0.95\linewidth]{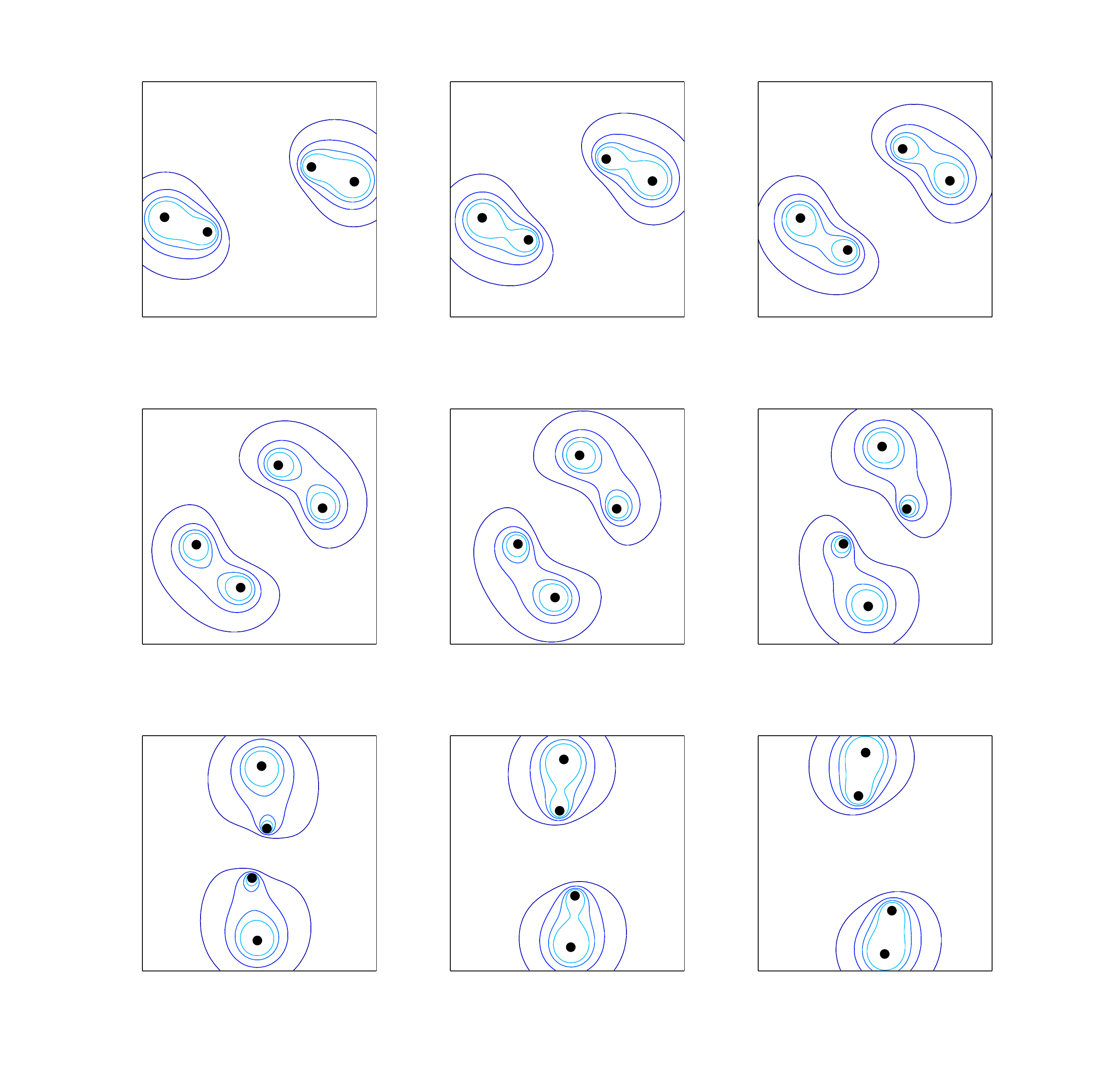}
\end{minipage}
\vspace{-0.5cm}
\caption{Geodesic for initial conditions $K/C=5+2\text{i}$, $\dot{K}/C=-0.042$ with step size $0.03$.  Tick 
marks are at every $950$ timesteps.  In this case the fundamental monopoles retain their separate 
identities.}\label{fig07}
\end{figure}

In complex coordinates the geodesic equations \eqref{geoeqs} are
\begin{equation*}
\eta\ddot{K}+(\partial_K\eta)\dot{K}^2\,=\,0
\end{equation*}
and its complex conjugate.  We write this as a system of coupled partial differential equations,
\begin{equation*}
\eta\dot{v}+(\partial_K\eta)v^2\,=\,0\qquad\qquad\dot{K}\,=\,v
\end{equation*}
and obtain $\partial_K\eta$ by differentiating the integrand of $\eta$ before performing the integral (this 
choice of ordering giving greater numerical precision),
\begin{equation*}
\partial_K\eta\,=\,\frac{1}{2}\,\int(\zeta^2-K/2)\left((\zeta^2-K/2)^2-C^2\right)^{-3/2}\left((\bar{\zeta}^2-\bar{K}/2)^2-C^2\right)^{-1/2}\,\rho\text{d}\rho\text{d}\theta,
\end{equation*}
which must again be integrated numerically.  Then, by specifying initial values of $K$ and $\dot{K}$, novel 
geodesics are integrated using a fourth order Runge-Kutta procedure.  Two such non-trivial geodesics are 
displayed in figs~\ref{fig06} and \ref{fig07}, which are to be compared with those of fig.~\ref{fig04}.  
It is worth noting that geodesics crossing the line segment $-2<K/C<2$ (fig.~\ref{fig06}) scatter by swapping 
constituents, otherwise there is glancing scattering and each fundamental monopole retains its identity 
(fig.~\ref{fig07}).  As was seen in fig.~\ref{fig04}, a geodesic meeting $K=\pm2C$ has two coincident 
spectral points, whose associated energy density vanishes.  There is numerical evidence that the only 
geodesic to cross these points is that with $K\in\mathbb{R}$.
\subsection{Zeroes on the Cylinder}\label{zeroes}
\begin{figure}
\centering
\includegraphics[width=12cm]{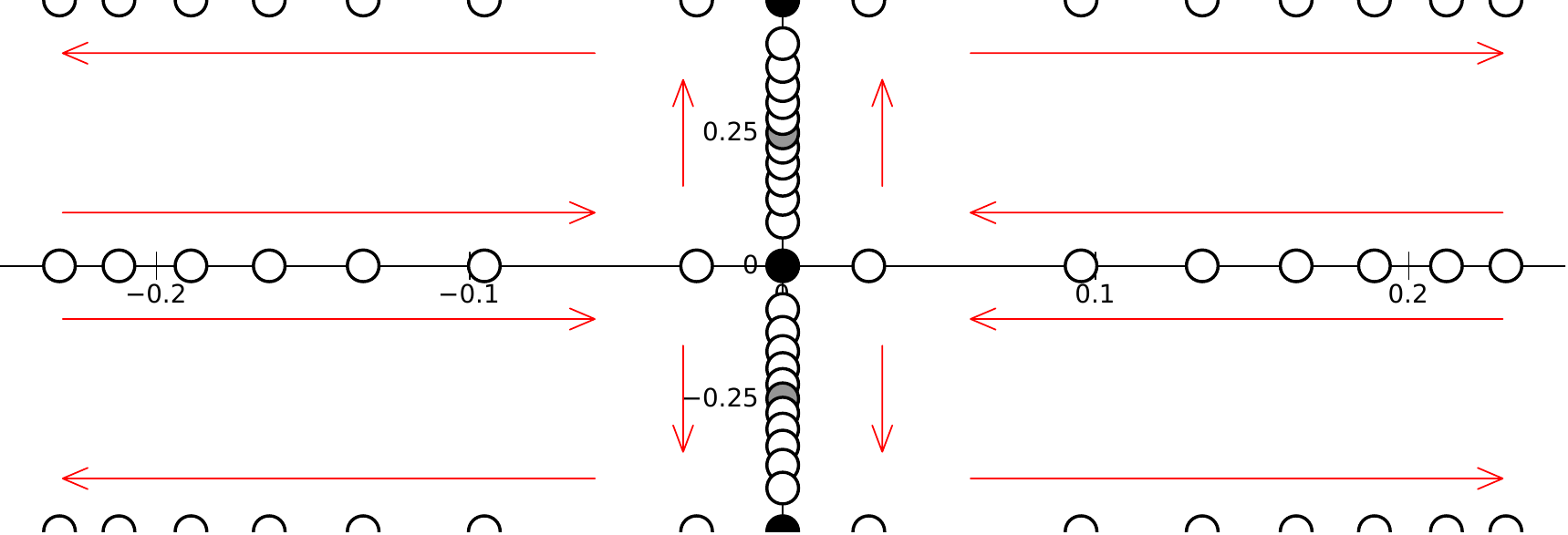}
\caption{Motion of zeroes on the Hitchin cylinder for $K\in\mathbb{R}$, where the top and bottom edges of the 
diagram are identified and the $z$ period is taken to be $\beta=2\pi$.  Arrows indicate the direction of $K$ 
increasing from $K/C=-4.5$, with spacing determined by the velocity using the metric 
\protect\eqref{metricintegral}.  The black dots are at $K/C=\pm2$ (note that in these cases the zeroes 
coincide), while the grey dots are at $K=0$.  Zeroes at the same $K$ are located at opposite points on the 
cylinder, obtained by reversing the signs of $r$ and $t$.}\label{fig08}
\end{figure}

\begin{figure}
\centering
\includegraphics[width=12cm]{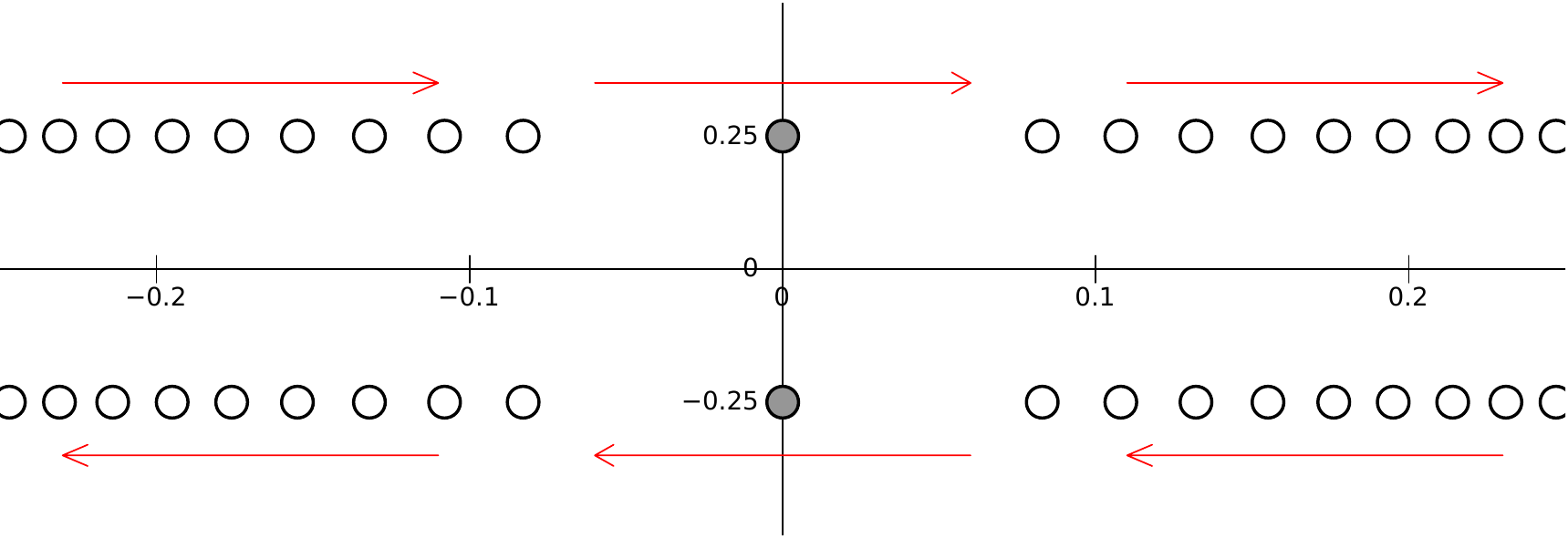}
\caption{The setup is the same as that of fig.~\ref{fig08}, this time with $K\in\text{i}\mathbb{R}$.  The 
arrow indicates the evolution with $\Im(K)$ increasing from $\Im(K)=-4.5$.}\label{fig09}
\end{figure}

Rewriting the spectral curve \eqref{speccharge2} as a polynomial in $\zeta$ and comparing with \eqref{hsc} we 
find
\begin{equation*}
\zeta^2-\left(C\cosh(\beta s)+K/2\right)\,=\,0\qquad\Rightarrow\qquad-\text{det}(\Phi)\,=\,C\cosh(\beta s)+K/2.\label{detcharge2}
\end{equation*}
The determinant of the Hitchin Higgs field has two zeroes whose locations on the cylinder depend on $K/C$.  
In section \ref{ch2} we will see that these values are of interest as they provide approximate locations for 
peaks in the gauge field $F_{s\bar{s}}$ on the Hitchin cylinder \eqref{hitchin}.  As $\cosh$ is an even 
function, the zeroes are always on opposite sides of the cylinder, at $\pm s_0$.  They are located on the 
circle $r=0$ if $-2\leq K/C\leq2$ and coincide at $s_0=\text{i}\pi/\beta,0$ if $K/C=2,-2$.  This suggests, as 
was noted in fig.~\ref{fig04}, that $K=0$ is a particularly symmetric case, for which the zeroes are at 
$\pm\text{i}\pi/2\beta$.  The motion of the zeroes corresponding to the geodesics with $K\in\mathbb{R}$ and 
$K\in\text{i}\mathbb{R}$ are shown in figs~\ref{fig08} and \ref{fig09}.  Other geodesics lead either to 
glancing scattering of the zeroes (if $K/C$ passes between $-2$ and $2$, such as in fig.~\ref{fig06}) or to 
them returning in the same direction they come in from (if $K/C$ does not cross the line segment $[-2,2]$, 
such as in fig.~\ref{fig07}).
\section{Charge 2 - Nahm Transform}\label{ch2}
The centered SU(2) charge $\bm{k}=(2)$ periodic monopole has four real moduli, two of which, as was seen in 
sections \ref{specdat} and \ref{charge2specaprox}, are encoded in the spectral curve and describe the 
relative positions and orientations of the fundamental monopoles in $\mathbb{R}^2$.  The remaining two moduli 
are expected to describe the relative phase and $z$ separation.  By considering the action of gauge 
transformations on the inverse Nahm operator (as defined in section \ref{nahm}) we will see that the two 
reduced moduli appearing in the spectral curve provide a geodesic submanifold of the full moduli space.  The 
one parameter families $K\in\mathbb{R}$ and $K\in\text{i}\mathbb{R}$ are studied, and we will find that the 
details of $z$ behaviour depend on our choice of solution of the Hitchin equations on the Hitchin cylinder.  
The work in this section is motivated by \cite{HW09,Har}, and it should be noted that the results are 
independent of the spectral approximation of section~\ref{ch2spap}.
\subsection{Hitchin Equations on the Cylinder}\label{hitchincharge2}
The Nahm data of interest are U(2) (or SU(2) if the monopole is centered) Hitchin fields ($\Phi,A$) 
\eqref{hitchin} on the Hitchin cylinder, with $\text{det}(\Phi)$ determined by the spectral curve as 
described in section \ref{zeroes}.  It is straightforward to show \cite{HW09} that the Hitchin equations can 
be solved (up to $\text{U}(1)$ gauge transformations) by\footnote{For the remainder of this section we make use of 
the Pauli matrices with conventions
\begin{equation*}
\sigma_1\,=\,\begin{pmatrix}0&1\\1&0\end{pmatrix}\qquad\qquad\sigma_2\,=\,\begin{pmatrix}0&-\text{i}\\\text{i}&0\end{pmatrix}\qquad\qquad\sigma_3\,=\,\begin{pmatrix}1&0\\0&-1\end{pmatrix}.
\end{equation*}}
\begin{equation}
\Phi\,=\,\begin{pmatrix}0&\mu_+\text{e}^{\psi/2}\\\mu_-\text{e}^{-\psi/2}&0\end{pmatrix}\qquad\qquad A_{\bar{s}}\,=\,a\sigma_3+\alpha\Phi\qquad\qquad A_s\,=\,-\bar{a}\sigma_3-\bar{\alpha}\Phi^\dag\label{pg}
\end{equation}
where
\begin{equation*}
-\text{det}(\Phi)\,=\,\mu_+\mu_-\,=\,C\cosh(\beta s)+K/2
\end{equation*}
and $a$, $\alpha$ and $\psi$ are functions of $(s,\bar{s})$ satisfying $4a=-\partial_{\bar{s}}\psi$,
\begin{equation}
\nabla^2\Re(\psi)\,=\,2(1+4|\alpha|^2)\left(|\mu_+|^2\text{e}^{\Re(\psi)}-|\mu_-|^2\text{e}^{-\Re(\psi)}\right)\label{psieq}
\end{equation}
and
\begin{equation}
\text{e}^{-\Re(\psi)/2}\,\partial_s\left(\alpha\mu_+\text{e}^{\Re(\psi)}\right)+\text{e}^{\Re(\psi)/2}\,\partial_{\bar{s}}\left(\bar{\alpha}\bar{\mu}_-\text{e}^{-\Re(\psi)}\right)\,=\,0,\label{alphaeq}
\end{equation}
with the imaginary part of $\psi$ chosen in such a way that $\Phi$ has the correct $t$-period.
\par It is clear that $\alpha=0$ allows \eqref{alphaeq} to hold trivially, and in the next subsection it will 
be seen that it in fact provides a two dimensional geodesic submanifold of the relative moduli space.  When 
this is the case, there are two fundamentally different solutions for $\Phi$ according to the allocation of 
the zeroes of $\text{det}(\Phi)$ between its two non-vanishing components:
\begin{itemize}
\item Harland's solution \cite{Har} places both zeroes in the same component,
\begin{equation*}
\mu_+\,=\,C\cosh(\beta s)+K/2\qquad\qquad\mu_-\,=\,1\label{muHar}
\end{equation*}
with $\Im(\psi)=0$.  We call this the `zeroes together' solution.
\item On the other hand, Harland \& Ward \cite{HW09} place one zero in each component of $\Phi$,
\begin{equation*}
\mu_\pm\,=\,\sqrt{C/2}\left(\text{e}^{\beta s/2}+\lambda^{\pm1}\text{e}^{-\beta s/2}\right)\qquad\text{where}\qquad2C\lambda^{\pm1}\,=\,K\pm\sqrt{K^2-4C^2}\label{muHW}
\end{equation*}
this time with $\Im(\psi)=\beta t$.  This is the `zeroes apart' solution.\footnote{In subsequent work 
\cite{MW,Mal} it has been found more meaningful to use the coordinate $\lambda$ rather than $K$.  Here, we 
retain the original notation for consistency with the published version.}
\end{itemize}
For $\alpha=0$ the Hitchin Higgs fields are thus of different matrix rank and there is no smooth gauge 
transformation between them.  As such, the `zeroes together' and `zeroes apart' solutions are disconnected 
two dimensional submanifolds of the moduli space.  It is expected that in the full four dimensional moduli 
space one can interpolate between the two cases.
\subsection{Symmetries}\label{fullsymms}
Once the Hitchin equations of section \ref{hitchincharge2} have been solved, one should apply the procedure 
of section \ref{nahm} to obtain the monopole fields.  This has been done numerically for the `zeroes apart' 
case \cite{HW09}.  Here, we consider symmetries of the Nahm transform by means of gauge transformations 
\eqref{psigauge}.  This is achieved by first looking for transformations of the Nahm data 
$(s;K)\mapsto(s';K')$ motivated by the findings of section \ref{charge2symmetries}, which should satisfy 
equations (\ref{psieq}, \ref{alphaeq}) and transform
\begin{IEEEeqnarray*}{rcl}
(\Phi,A)(s;K)&\,\mapsto\,&(\Phi',A')(s';K')\\
(\Delta,\Psi)(s;\zeta',z';K)&\,\mapsto\,&(\Delta,\Psi)(s';\zeta',z';K')\,=\,(\Delta',\Psi')(s;\zeta',z';K),
\end{IEEEeqnarray*}
(where we use $'$ to denote fields valued in the transformed Hitchin coordinates) and then searching for a 
gauge transformation $U$ and a transformation $(\zeta,z)\mapsto(\zeta',z')$ of the monopole coordinates which 
express $\Delta'$ in terms of $\Delta$, in such a way that the resulting monopole fields are gauge equivalent 
to the original monopole fields, but evaluated at the new coordinates, $(\zeta',z')$.  We recall from equation 
\eqref{psigauge} in section \ref{nahm} that $U$ acts as
\begin{IEEEeqnarray*}{rcl}
\Delta'(s;\zeta',z';K)&\,=\,&U^{-1}(s)\Delta(s;\zeta',z';K)U(s)\\
\Psi'(s;\zeta',z';K)&\,=\,&U^{-1}(s)\Psi(s;\zeta',z';K),
\end{IEEEeqnarray*}
and we assume it can be written in block form as $U=h\otimes g$, where $h$ is a constant $2\times2$ matrix 
serving to permute the entries of $\Delta$.  The matrix $g$ acts as a gauge transformation on the Hitchin 
fields and is required to be strictly periodic in $t$, such that $\Phi$ and the $t$-holonomy of $A$ are well 
defined.  For completeness, we recall the Nahm operator \eqref{invnahm} in the $\bm{k}=(2)$ case,
\begin{equation}
\Delta\,=\,\begin{pmatrix}\bm{1}_2\otimes(2\partial_{\bar{s}}-z)+2A_{\bar{s}}&\bm{1}_2\otimes\zeta-\Phi\\\bm{1}_2\otimes\bar{\zeta}-\Phi^\dag&\bm{1}_2\otimes(2\partial_s+z)+2A_s\end{pmatrix}_.\label{nahmopcharge2}
\end{equation}
A study of the geodesic with $\alpha=0$, $K\in\mathbb{R}$ and the symmetry $K\mapsto-K$ was carried out in 
\cite{Har,HW09}.  Here we summarise the results and give evidence that $K\in\text{i}\mathbb{R}$ is also a 
geodesic.
\subsubsection*{$\underline{z\mapsto z+\beta}$}
To illustrate the process, we note the Hitchin fields are unchanged under the joint action of 
$U=\text{e}^{-\text{i}\beta t}\bm{1}_4$ and $(\zeta,z)\mapsto(\zeta,z+\beta)$, indicating the monopole fields 
are unchanged by a period shift.
\subsubsection*{$\underline{\alpha=0}$}
Again keeping $s$ and $K$ unchanged, we take $U=\sigma_3\otimes\bm{1}_2$ and $(\zeta,z)\mapsto(-\zeta,z)$.  
As long as $\alpha=0$ the Hitchin fields become $(\Phi,A)\mapsto(-\Phi,A)$, so that 
$\Psi_\pm\mapsto\pm\Psi_\pm$ and the monopole fields are thus invariant under a rotation by $\pi$ around the 
$z$-axis.  This justifies our assumption throughout section \ref{ch2spap} that $\alpha=0$ is a geodesic 
submanifold, and we will keep $\alpha=0$ from now on, noting that this simplifies the Hitchin gauge potential 
$A$ and that the Hitchin equation \eqref{alphaeq} is automatically satisfied.  Symmetries with $\alpha\neq0$ 
are considered by the author in his PhD thesis \cite{Mal14}.
\subsubsection*{$\underline{K\in\mathbb{R}}$}
Transforming $(s;K)\mapsto(\bar{s};\bar{K})$ gives 
$(\Phi',A'_{\bar{s}},A'_s)=(\sigma_1\Phi^\dag\sigma_1,-A_s,-A_{\bar{s}})$.  We then take 
$U=\sigma_1\otimes\sigma_1$ and $(\zeta,z)\mapsto(\bar{\zeta},-z)$.  As was found in section 
\ref{charge2symmetries}, $K\in\mathbb{R}$ is a geodesic submanifold, and the monopole fields are invariant 
under a joint reflection in the $x$-axis and the plane $z=0$ (or $z=\frac{\beta}{2}$).  This conclusion can 
be drawn for both solutions considered in section \ref{hitchincharge2}.  The calculation for the `zeroes 
together' case is given in more detail in appendix \ref{appen}, which serves to illustrate the procedure 
for the remaining cases.
\subsubsection*{$\underline{K\in\mathrm{i}\,\mathbb{R}}$ `zeroes together'}
The transformation $(s;K)\mapsto(\bar{s}+\text{i}\frac{\pi}{\beta};-\bar{K})$ gives 
$(\Phi',A'_{\bar{s}},A'_s)=(-\text{i}g^{-1}\Phi^\dag g,-A_s,-A_{\bar{s}})$, where 
$g=\frac{\text{i}}{\sqrt{2}}(\sigma_1+\sigma_2)$.  Now we must take $h=\frac{1}{\sqrt{2}}(\sigma_1+\sigma_2)$ 
and $(\zeta,z)\mapsto(\text{i}\bar{\zeta},-z)$.
\subsubsection*{$\underline{K\in\mathrm{i}\,\mathbb{R}}$ `zeroes apart'}
Here the same map $(s;K)\mapsto(\bar{s}+\text{i}\frac{\pi}{\beta};-\bar{K})$ gives 
$(\Phi',A'_{\bar{s}},A'_s)=(\text{i}g^{-1}\Phi^\dag g,A_s+\frac{\beta}{4}\sigma_3,A_{\bar{s}}-\frac{\beta}{4}\sigma_3)$, 
with
\begin{equation*}
g\,=\,\begin{pmatrix}\text{e}^{\text{i}\beta t}&0\\0&\text{i}\end{pmatrix},
\end{equation*}
and we take $h=\frac{1}{\sqrt{2}}(\sigma_1-\sigma_2)$ and 
$(\zeta,z)\mapsto(\text{i}\bar{\zeta},\frac{\beta}{2}-z)$.
\subsubsection*{$\underline{K\mapsto-K}$ `zeroes together'}
We take $(s;K)\mapsto(s+\text{i}\frac{\pi}{\beta},-K)$, so $(\Phi',A')=(\text{i}g^{-1}\Phi g,A)$ where 
$g=\text{e}^{-\text{i}\pi\sigma_3/4}$, $h=\text{e}^{-\text{i}\pi\sigma_3/4}$ and 
$(\zeta,z)\mapsto(\text{i}\zeta,z)$.
\subsubsection*{$\underline{K\mapsto-K}$ `zeroes apart'}
Finally, with the same $(s';K')$ as the previous case, 
$(\Phi',A'_{\bar{s}},A'_s)=(\text{i}g^{-1}\Phi g,\frac{\beta}{4}\sigma_3-A_{\bar{s}},-\frac{\beta}{4}\sigma_3-A_s)$, 
with
\begin{equation*}
g\,=\,\begin{pmatrix}0&\text{e}^{\text{i}\beta t+\text{i}\pi/4}\\\text{e}^{-\text{i}\pi/4}&0\end{pmatrix},
\end{equation*}
such that $h=\text{e}^{-\text{i}\pi\sigma_3/4}$ and $(\zeta,z)\mapsto(\text{i}\zeta,z+\frac{\beta}{2})$.
\par Although the only currently available numerical solution for the monopole fields is the $K\in\mathbb{R}$ 
geodesic in the `zeroes apart' case \cite{HW09}, the above results show that all four possibilities undergo 
right-angled scattering, with a configuration of enhanced symmetry at $K=0$.  Nevertheless, in the `zeroes 
together' solution, scattering occurs in a plane of constant $z$, while in the `zeroes apart' solution, the 
incoming and outgoing chains are shifted by half a period.  The two situations can be visualised as chains of 
small monopoles (though it should be noted this is no longer the {\it r\'egime} in which we expect the 
spectral approximation to be valid), which scatter at an angle of $\pi/2$.  Such a scattering process may 
occur in the plane or along $z$, in which case there will be a second scattering when the fundamental 
monopoles meet those of the adjacent periods and then separate at right angles to the incoming chains.
\par The symmetric configuration with $K=0$ is seen to survive under the arguments given above.  It has also 
been noted \cite{HW09} that when $K=\pm2C$ the periodic monopole resembles a unit charge periodic monopole of 
halved period.  This is consistent with the observation that the result of the spectral approximation for 
this case (fig.~\ref{fig04}) resembles the result for charge $1$, fig.~\ref{fig02}.
\subsection{Lumps on the Cylinder}\label{lumpscyl}
\begin{figure}
\begin{minipage}{0.495\textwidth}
\centering
\includegraphics[width=\linewidth]{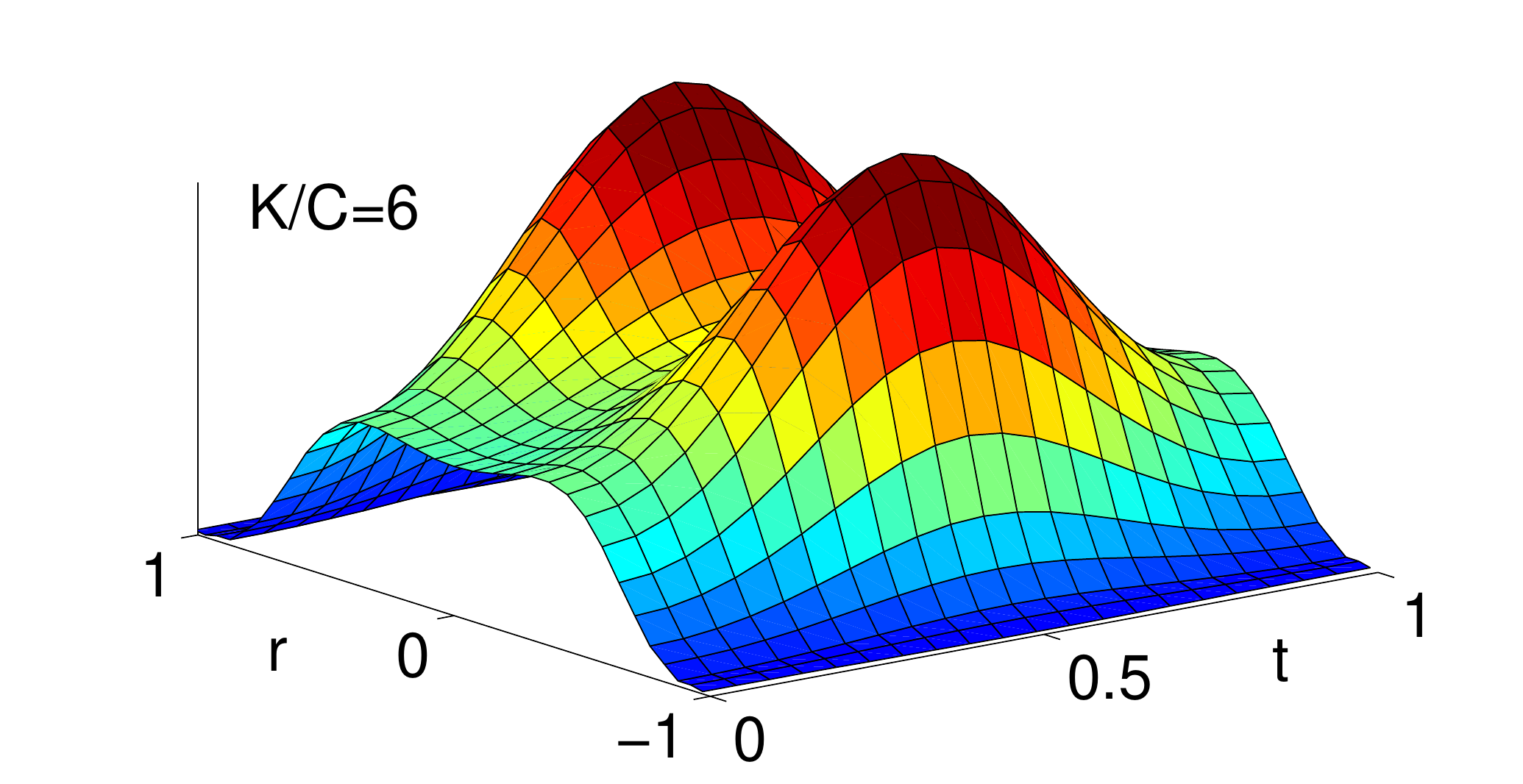}
\includegraphics[width=\linewidth]{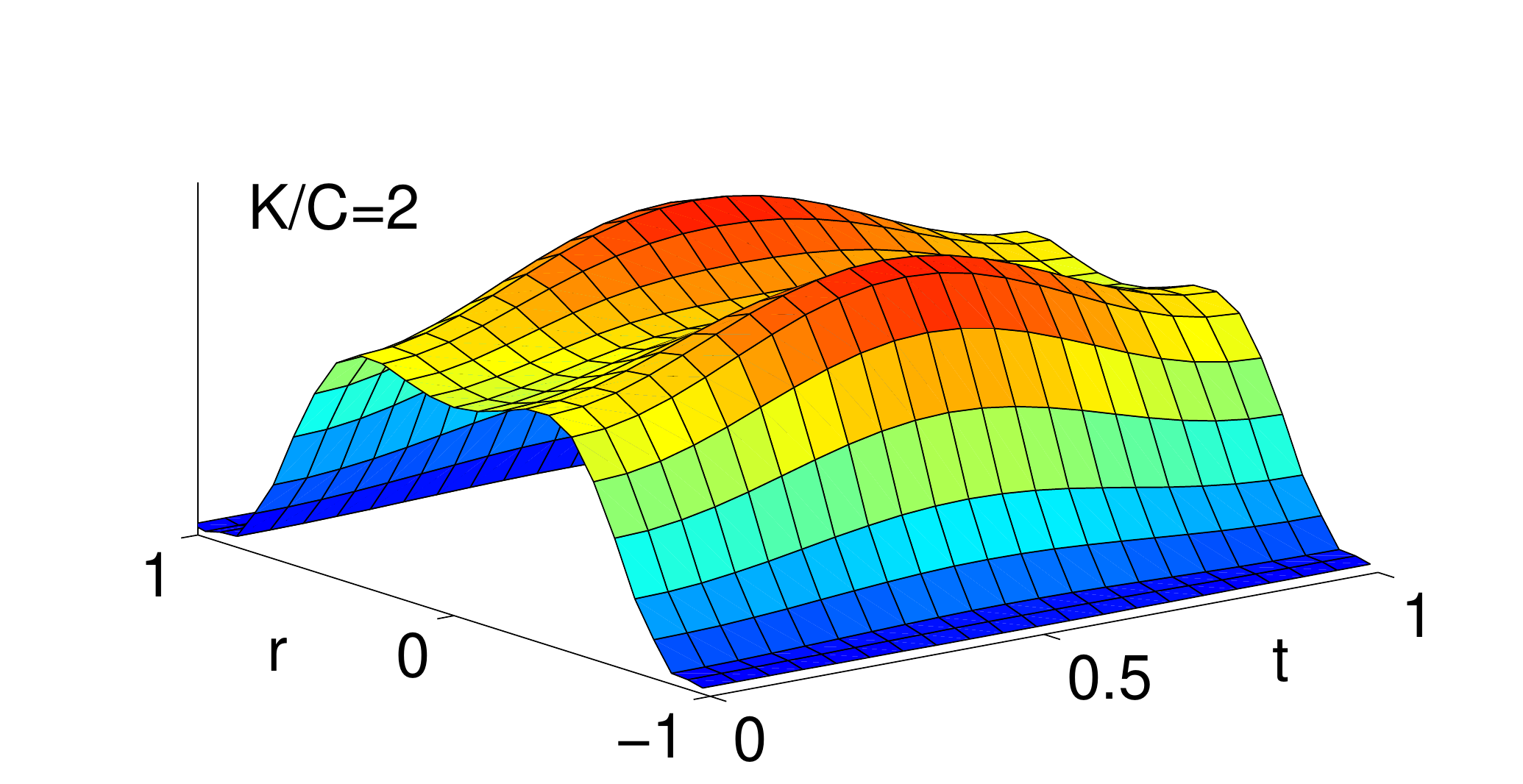}
\includegraphics[width=\linewidth]{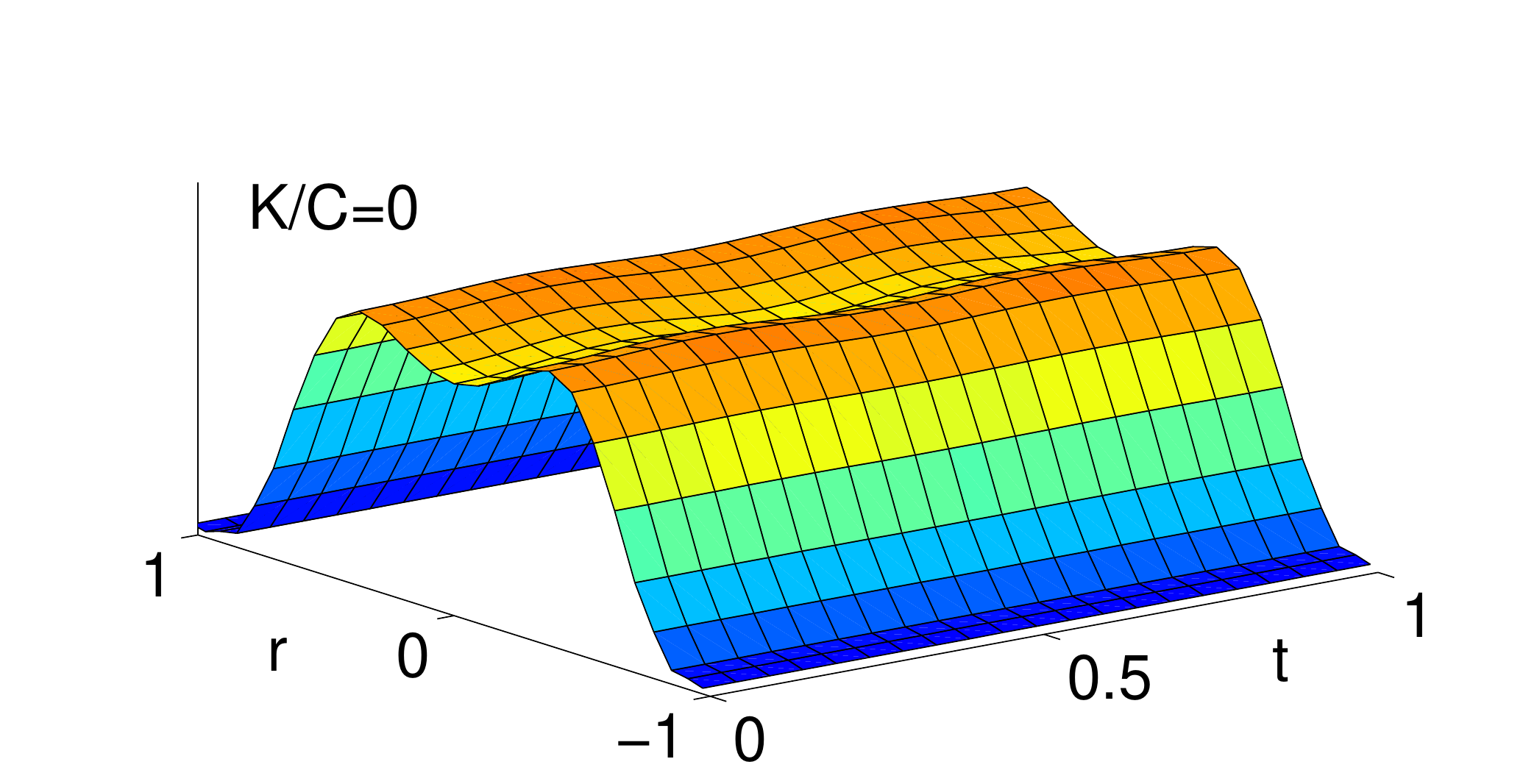}
\end{minipage}
\begin{minipage}{0.495\textwidth}
\centering
\includegraphics[width=\linewidth]{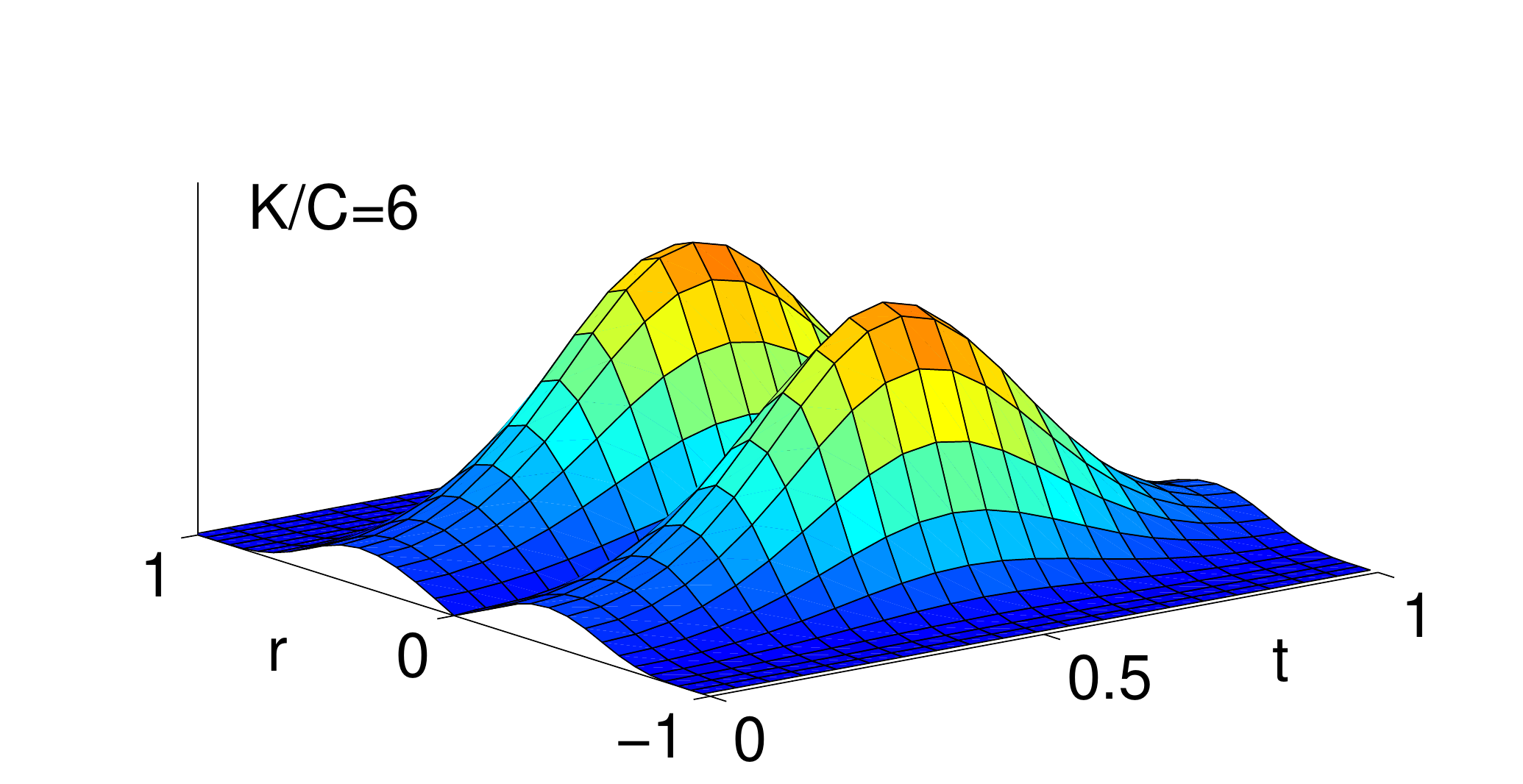}
\includegraphics[width=\linewidth]{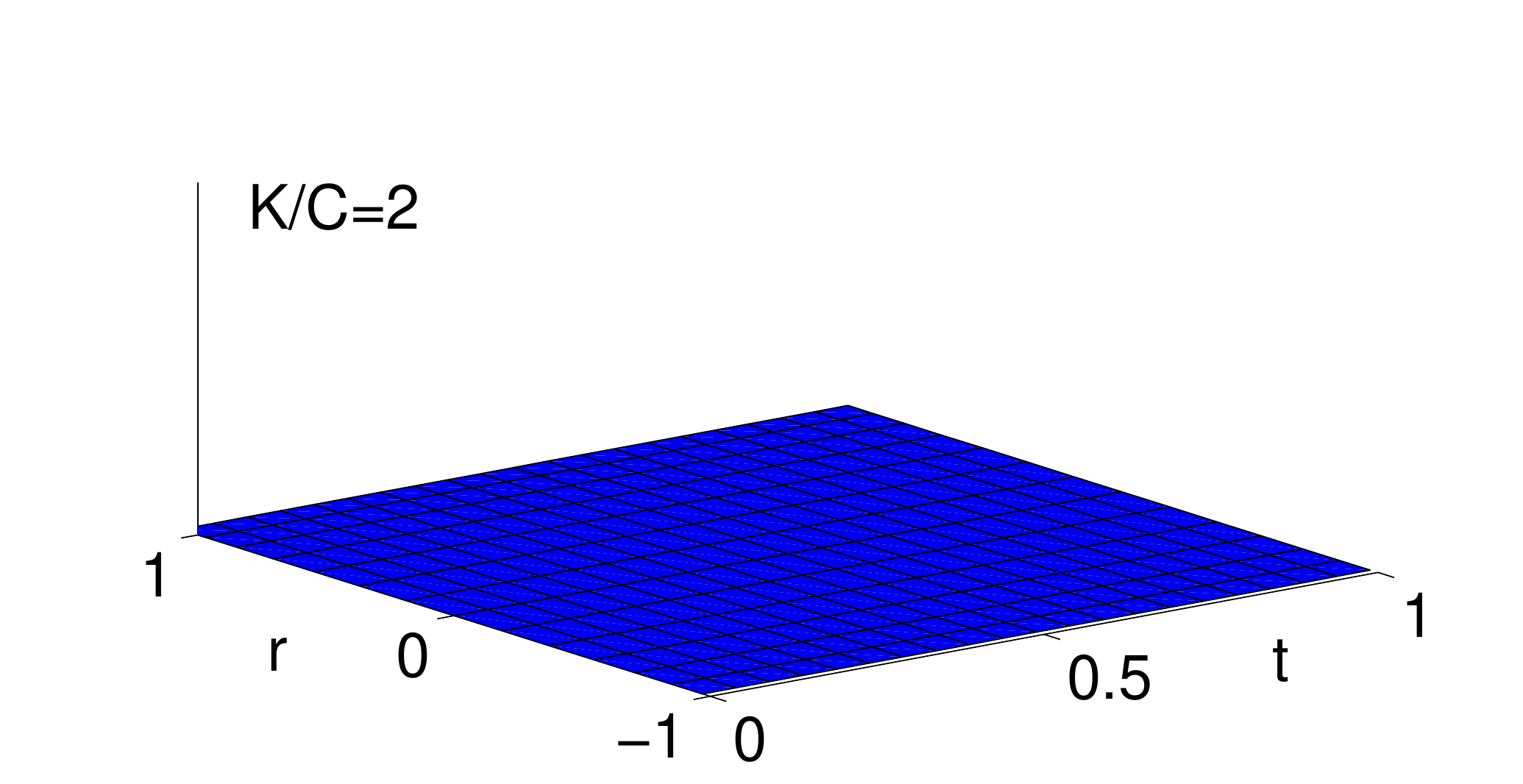}
\includegraphics[width=\linewidth]{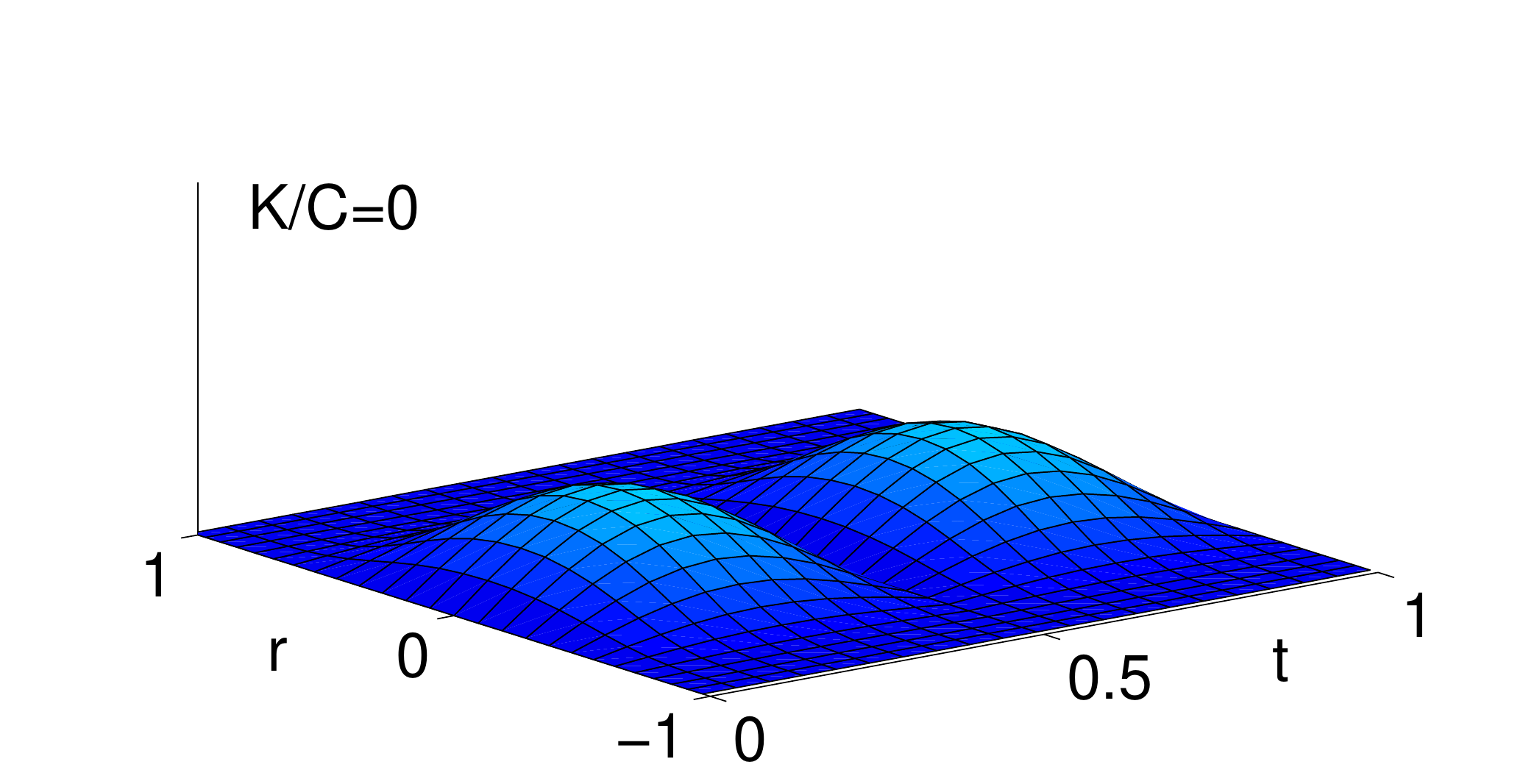}
\end{minipage}
\caption{Lumps in $|F_{s\bar{s}}|$ for various values of $K$ in the `zeroes together' solution (left) and 
`zeroes apart' (right) for $C=1$, $\beta=2\pi$, using the same vertical scale throughout.  The positions of 
the lumps should be compared with the positions of the zeroes of $\text{det}(\Phi)$, as indicated in 
figs~\protect\ref{fig08} and \protect\ref{fig09}.  It should also be noted that the lumps are of different 
sign in each case.}\label{fig10}
\end{figure}

Working with $\alpha=0$, the remaining Hitchin equation \eqref{psieq} can be solved numerically using a 
relaxation method, \cite{HW09}.  Fig.~\ref{fig10} displays the value of the flux 
$|F_{s\bar{s}}|=\frac{1}{8}|\nabla^2\psi|$, for which the general characteristics can be deduced from 
\eqref{psieq}.  In particular, in the `zeroes apart' case the lumps annihilate at $K=\pm2C$, when both 
$\mu_+$ and $\mu_-$ vanish.  On the other hand, in the `zeroes together' solution the lumps do not vanish, 
but reach a minimum size at $K=0$.
\par Numerically, a dependence on $C$ is also observed, with two limiting cases.  For small monopole size $C$ 
the lumps lose $t$-dependence and become Nahm data on a line segment.  However, at large $C$ (which is the 
case of interest in section~\ref{ch2spap} of this paper) the lumps become sharply peaked and \eqref{psieq} 
is solved by setting both sides to zero.  It is in the latter case that the spectral approximation improves 
in accuracy, and that the positions of the lumps are found to most closely track the zeroes of 
$\text{det}(\Phi)$ shown in figs~\ref{fig08} and \ref{fig09}.
\section{SU(3) Periodic Monopoles}\label{su3pdcmon}
Monopoles in $\mathbb{R}^3$ have been considered for higher rank gauge groups by various authors 
\cite{MS07,Wei79,LWY96,Shn05}.  In this section we apply the results of the spectral approximation to the 
SU(3) periodic monopole and consider the basic properties for $\bm{k}=(1,1)$ and $\bm{k}=(2,1)$, which have 
two and four reduced relative moduli, respectively.
\par Following the arguments of section \ref{monopoledata}, the SU(3) periodic monopole has spectral curve 
\eqref{msc}
\begin{equation*}
w^3+P_{1,k_1}(\zeta)w^2+P_{2,k_2}(\zeta)w-1\,=\,0\label{spec}\qquad\text{where}\qquad P_{i,k_i}(\zeta)\,=\,a_{i,k_i}\zeta^{k_i}+\ldots+a_{i,1}\zeta+a_{i,0}.
\end{equation*}
As discussed in sections \ref{monopoledata} and \ref{specdat}, we take $k_1\geq k_2$.  The root diagram was shown in 
fig.~\ref{fig01}.  Our procedure will be to express the coefficients of $P_{i,k_i}(\zeta)$ in terms of the 
boundary data (\ref{asympfields}, \ref{asympholo}) and hence to determine the positions of spectral points 
from the discriminant $\mathcal{D}_{(k_1,k_2)}$.  In analogy with section \ref{spcu}, we are interested in the 
eigenvalues of the holonomy $V$.  This manipulation is performed numerically to give three eigenvalues 
$w_i=\text{exp}(\beta (r_i+\text{i}t_i))$ from which $\hat{\Phi}\propto\text{diag}(r_1,r_2,r_3)$ and the 
quantities of interest are\footnote{In the SU(2) case (section \protect\ref{spap} and 
fig.~\protect\ref{fig02}) a similar calculation gave $\hat{\Phi}=\text{i}r_0\sigma_3$, 
$\mathcal{E}\propto\nabla^2r_0^2$ and $\text{disc.}=4r_0^2$.}
\begin{equation*}
\mathcal{E}\,\propto\,\nabla^2\left(r_1^2+r_2^2+r_3^2\right),\qquad\qquad\text{discriminant}\,=\,(r_1-r_2)^2(r_2-r_3)^2(r_3-r_1)^2.
\end{equation*}
\subsection{Trivial Embedding}\label{trivialembedding}
The $\bm{k}=(1,1)$ spectral curve has coefficients
\begin{equation}
\bm{k}\,=\,(1,1)\qquad\left\lbrace\begin{array}{c c c}a_{1,1}\,=\,-\text{e}^{\mathfrak{v}_1} &&a_{1,0}\,=\,-(\mu_1\text{e}^{\mathfrak{v}_1}+\text{e}^{\mathfrak{v}_2})\\
a_{2,1}\,=\,\text{e}^{\mathfrak{v}_1+\mathfrak{v}_2}&&a_{2,0}\,=\,(\mu_1+\mu_2)\text{e}^{\mathfrak{v}_1+\mathfrak{v}_2}+\text{e}^{-\mathfrak{v}_2},\end{array}\right.\label{k11coeffs}
\end{equation}
and discriminant
\begin{equation*}
\mathcal{D}_{(1,1)}\,=\,a_{1,1}^2a_{2,1}^2\zeta^4+2\left(a_{1,1}a_{2,1}(a_{1,1}a_{2,0}+a_{1,0}a_{2,1})+2(a_{1,1}^3-a_{2,1}^3)\right)\zeta^3+\ldots,
\end{equation*}
such that the spectral points are centered if the $\zeta^3$ term vanishes,
\begin{equation*}
(2\mu_1+\mu_2)\text{e}^{\mathfrak{v}_1+2\mathfrak{v}_2}\,=\,\text{e}^{3\mathfrak{v}_2}+1.
\end{equation*}
As noted in section \ref{specdat}, the fact $K$ is repeated means the Nahm data will have a singularity at 
finite $|r|$.  As we are working with SU(3) the determinant will have three zeroes.
\par If $\mathfrak{v}_2=0$ and $\mu_2=0$ (such that the centering condition becomes 
$\mu_1\text{e}^{\mathfrak{v}_1}=1$) the monopole is an SU(2) monopole embedded along the root 
$\bm{\beta}_3^\ast=-\bm{\beta}_1^\ast-\bm{\beta}_2^\ast$.  This allows the spectral curve to be factorised,
\begin{equation*}
(w-1)\left(w^2-(\text{e}^{\mathfrak{v}_1}\zeta+1)w+1\right)\,=\,0.
\end{equation*}
In this limit, three of the spectral points coincide and, as expected, the monopole fields resemble those 
of an SU(2) monopole with $\bm{k}=(1)$.
\subsubsection*{$\underline{\mathfrak{v}_2\neq0}$}
\begin{figure}
\centering
\includegraphics[width=0.8\textwidth]{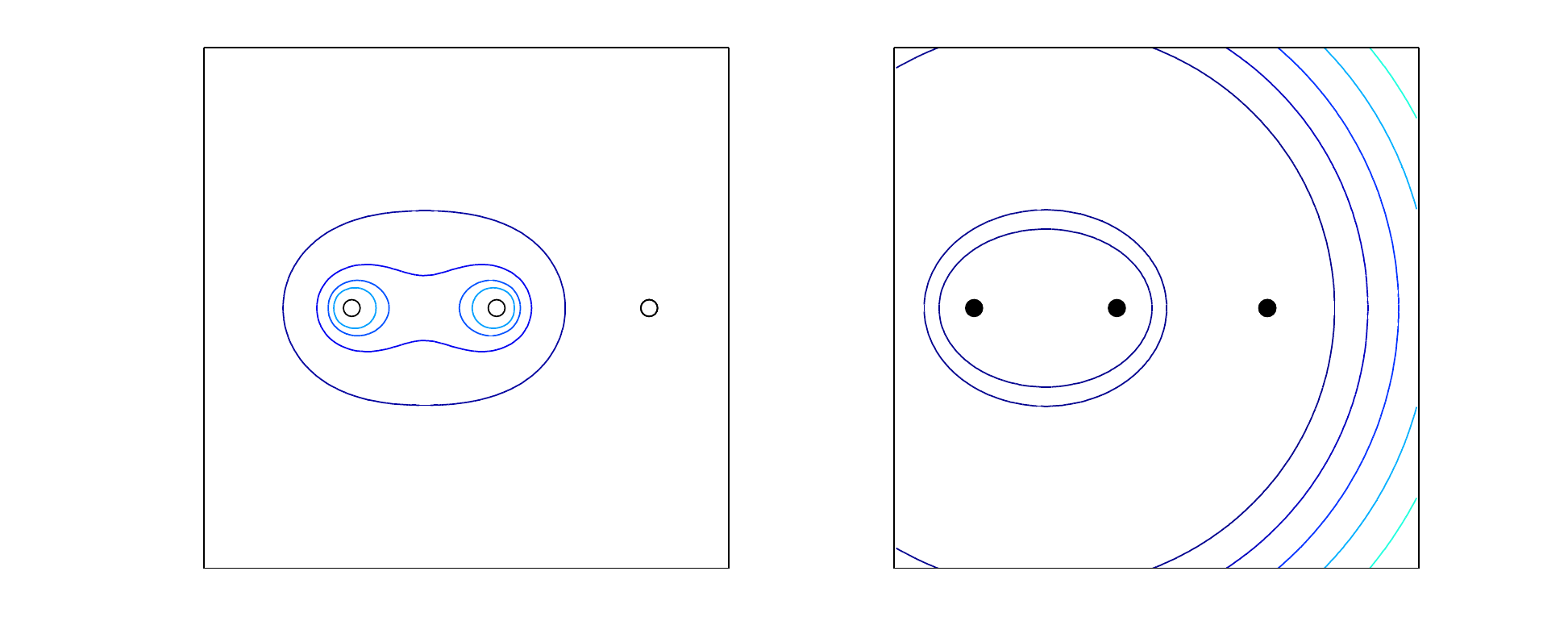}
\caption{Deformations of the $\bm{k}=(1,1)$ monopole by changing $\mathfrak{v}_2$ away from zero.  Here 
$\mathfrak{v}_2=1.2$.  On the left is plotted the energy density and on the right the discriminant of 
$\hat{\Phi}$.  There is no energy density associated with the coincident spectral points on the 
right.  The discriminant vanishes on a line joining the spectral points on the left, and on a circle passing 
through the double spectral point and surrounding the other two.}\label{fig11}
\end{figure}

We deform away from the SU(2) embedding by changing the boundary conditions to allow non-zero 
$\mathfrak{v}_2$.  The spectral curve again factorises, and centering identifies
\begin{equation*}
a_{1,0}\,=\,-\tfrac{1}{2}\left(3\text{e}^{\mathfrak{v}_2}+\text{e}^{-2\mathfrak{v}_2}\right)\qquad\qquad a_{2,0}\,=\,\tfrac{1}{2}\left(\text{e}^{2\mathfrak{v}_2}+3\text{e}^{-\mathfrak{v}_2}\right).
\end{equation*}
with $a_{1,1}$ and $a_{2,1}$ as in \eqref{k11coeffs}.  The situation is shown in fig.~\ref{fig11}.  In 
the Nahm picture, the Higgs field has a simple pole at $s=\mathfrak{v}_2/\beta$.  For $\mu_2=0$ one of the 
zeroes coincides with the pole, giving the two zeroes characteristic of SU(2) solutions.
\subsubsection*{$\underline{\mu_2\neq0}$}
\begin{figure}
\centering
\includegraphics[width=0.8\textwidth]{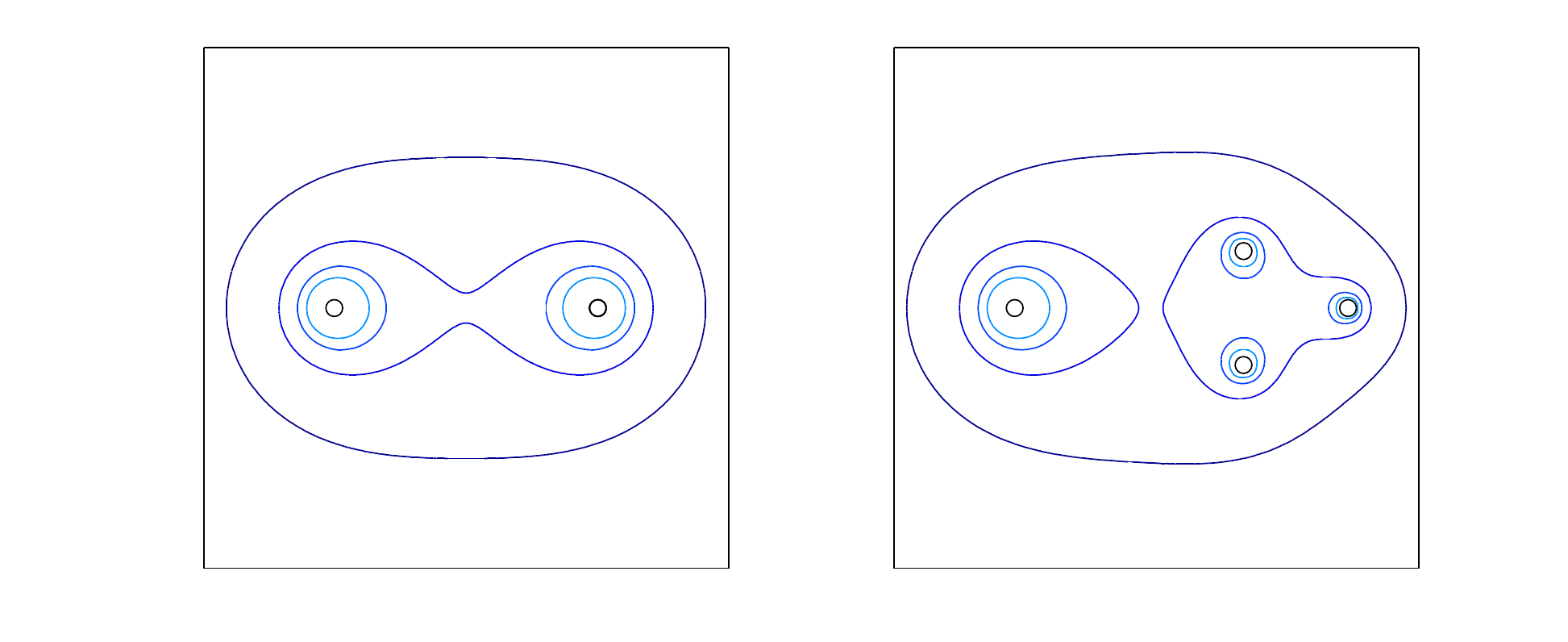}
\caption{Deformations of the $\bm{k}=(1,1)$ monopole with $\mathfrak{v}_2=0$.  On the left are contours of 
energy density for $\mu_1\text{e}^{\mathfrak{v}_1}=1$.  On the right, for 
$\mu_1\text{e}^{\mathfrak{v}_1}=1.2$.  For these examples, the discriminant pairs up the spectral 
points on the horizontal axis.  The line of zero discriminant joining the other two points is found to wrap 
around the left hand spectral point.}\label{fig12}
\end{figure}

In a similar way, we can fix the boundary conditions to $\mathfrak{v}_2=0$ and allow the moduli $\mu_1$ and 
$\mu_2$ to vary in such a way that the spectral points remain centered.  The coefficients in 
\eqref{k11coeffs} become
\begin{equation*}
a_{1,1}\,=\,-\text{e}^{\mathfrak{v}_1}\qquad\qquad a_{1,0}\,=\,-(1+\mu_1\text{e}^{\mathfrak{v}_1})\qquad\qquad a_{2,1}\,=\,\text{e}^{\mathfrak{v}_1}\qquad\qquad a_{2,0}\,=\,3-\mu_1\text{e}^{\mathfrak{v}_1}.
\end{equation*}
Varying $\mu_1$ separates the three coincident spectral points and introduces a second fundamental monopole, 
as shown in fig.~\ref{fig12}.
\subsection{Minimal Symmetry Breaking}\label{minsb}
\begin{figure}
\centering
\includegraphics[width=0.8\textwidth]{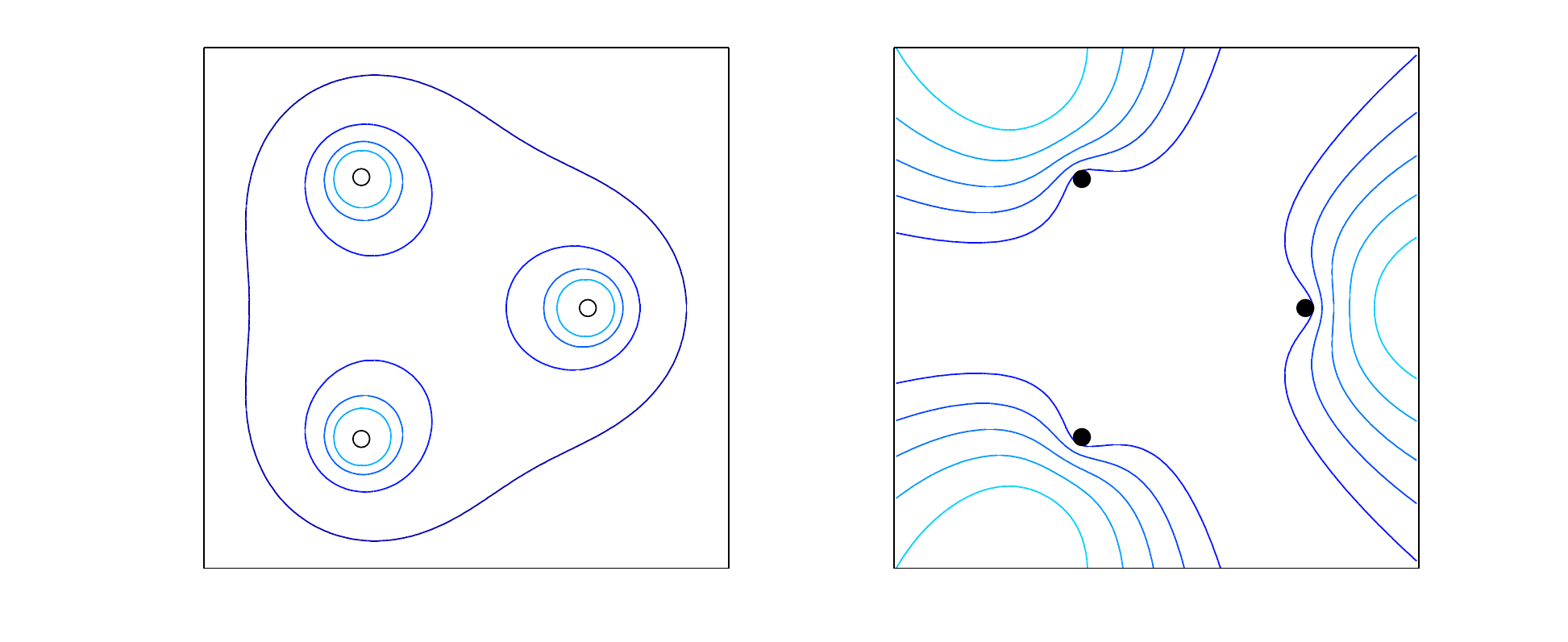}
\caption{$\mathbb{Z}_3$-symmetric $\bm{k}=(2,1)$ periodic monopole with spectral curve 
$w^3-\zeta^2w^2+2\zeta w-1=0$.  Energy density on the left and the discriminant of $\hat{\Phi}$ on the 
right.}\label{fig13}
\end{figure}

The $\bm{k}=(2,1)$ spectral curve has
\begin{equation*}
\bm{k}\,=\,(2,1)\qquad\left\lbrace\begin{array}{c c c}a_{1,2}\,=\,-\text{e}^{\mathfrak{v}_1} &&a_{1,1}\,=\,-\mu_1\text{e}^{\mathfrak{v}_1}\\
a_{2,1}\,=\,\text{e}^{\mathfrak{v}_1+\mathfrak{v}_2}+\text{e}^{-\mathfrak{v}_2}&&a_{2,0}\,=\,(\mu_1+\mu_2)\text{e}^{\mathfrak{v}_1+\mathfrak{v}_2}-\mu_2\text{e}^{-\mathfrak{v}_2},\end{array}\right.\label{k21coeffs}
\end{equation*}
and discriminant
\begin{equation*}
\mathcal{D}_{(2,1)}\,=\,a_{1,2}^2\left(a_{2,1}^2+4a_{1,2}\right)\zeta^6+2a_{1,2}\left(a_{1,2}a_{2,1}a_{2,0}+a_{1,1}a_{2,1}^2+6a_{1,1}a_{1,2}\right)\zeta^5+\ldots
\end{equation*}
and the remaining coefficient, $a_{1,0}$, is to be considered a modulus.  In this case, two of the $\ell_i$ 
are repeated, allowing minimal symmetry breaking if 
$\bm{\mathfrak{v}}=(2\mathfrak{v},-\mathfrak{v},-\mathfrak{v})$, for which centering implies that
\begin{equation*}
a_{1,2}\,=\,-\text{e}^{2\mathfrak{v}}\qquad\qquad a_{1,1}\,=\,-\mu_1\text{e}^{2\mathfrak{v}}\qquad\qquad a_{2,1}\,=\,2\text{e}^{\mathfrak{v}}\qquad\qquad a_{2,0}\,=\,\mu_1\text{e}^{\mathfrak{v}}.
\end{equation*}
In fact, this condition is equivalent to the coefficient of $\zeta^6$ in $\mathcal{D}_{(2,1)}$ vanishing, which 
was not a possibility for the SU(2) or $\bm{k}=(1,1)$ cases considered so far.  The coefficient of $\zeta^4$ 
also vanishes if we set $P_{1,2}=-\frac{1}{4}P_{2,1}^2$, such that three of the spectral points are sent to 
infinity.  This leaves $\mu_1$ as a complex modulus, and a symmetric configuration is obtained by taking 
$\mu_1=0$, such that the coefficients of $\zeta^2$ and $\zeta$ also vanish, fig.~\ref{fig13}.
\subsubsection*{$\underline{\mathfrak{v}_2\neq\mathfrak{v}_3}$}
\begin{figure}
\centering
\includegraphics[width=6cm]{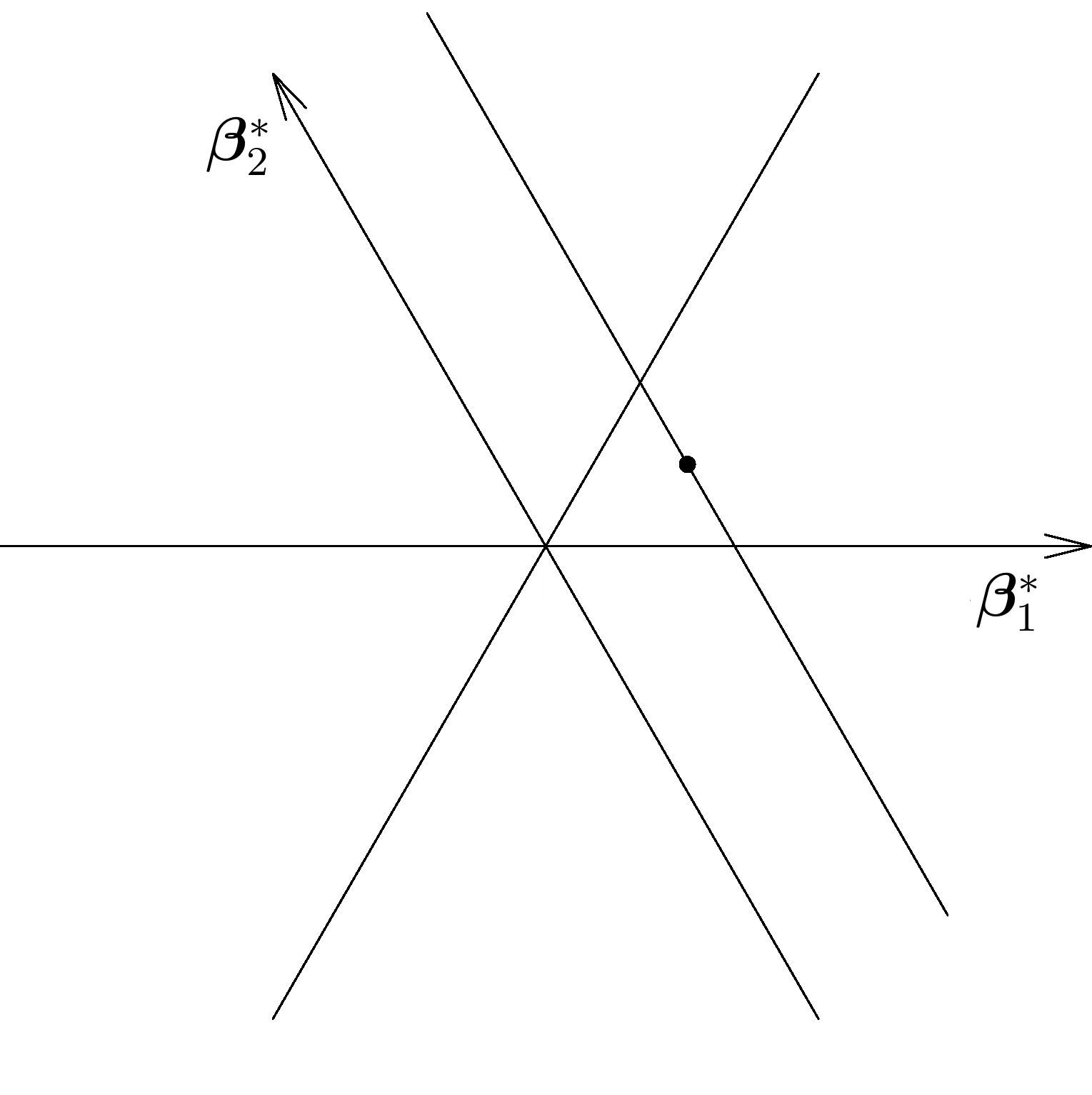}
\caption{Deformation of the subleading term.  Starting from the shaded point we deform parallel to 
$\bm{\beta}_2^\ast$.}\label{fig14}
\end{figure}

\begin{figure}
\centering
\includegraphics[width=0.8\textwidth]{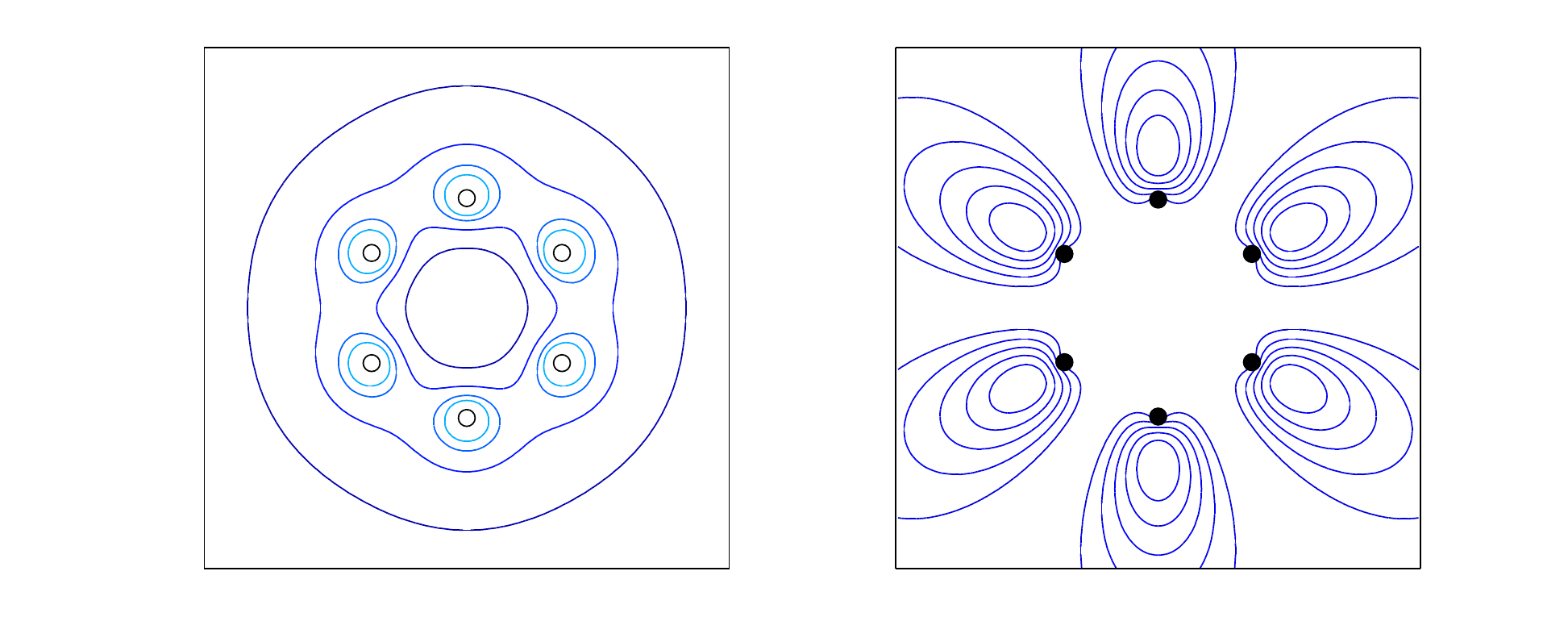}
\caption{$\mathbb{Z}_6$-symmetric $\bm{k}=(2,1)$ periodic monopole with spectral curve $w^3-\zeta^2w^2-1=0$.  
On the left is plotted the energy density and on the right the discriminant of $\hat{\Phi}$.}\label{fig15}
\end{figure}

Following \cite{War82} we deform by adding to $\bm{\mathfrak{v}}$ a constant diagonal term 
$\delta\bm{\beta}_2^\ast$ for some complex $\delta$ (we can rearrange the entries such that $\Re(\delta)\geq0$), 
fig.~\ref{fig14}.  The total energy \eqref{energy} is unchanged, but there is a different pattern of 
symmetry breaking.  Explicitly, $a_{1,2}$ and $a_{1,1}$ are unaltered, while
\begin{equation*}
a_{2,1}\,=\,2\text{e}^{\mathfrak{v}}\cosh(\delta)\qquad\qquad a_{2,0}\,=\,\text{e}^{\mathfrak{v}}\left(\mu_1\text{e}^\delta+2\mu_2\sinh(\delta)\right).
\end{equation*}
Such deformations have the effect of moving the three remaining spectral points in from infinity.  A 
particularly symmetric example, with $\delta=\text{i}\pi/2$, is shown in fig.~\ref{fig15}.
\par The $\bm{k}=(2,1)$ Nahm data is of rank $2$, smooth, and has three zeroes.  For the spectral curve 
$w^3-\zeta^2w^2+2a\zeta w-1=0$ relevant to both the cases considered above, the Hitchin Higgs fields have
\begin{equation*}
\text{tr}(\Phi)\,=\,2aw^{-1}\qquad\qquad-\text{det}(\Phi)\,=\,w-w^{-2}.
\end{equation*}
The determinant has zeroes at $\beta s=0,\pm2\text{i}\pi/3$.  This is reminiscent of the fact that the most 
symmetric $\bm{k}=(2)$ configurations were found to have zeroes located symmetrically on the Hitchin cylinder 
(fig.~\ref{fig08}).
\subsection{Speculative Geodesic}
In section \ref{fullsymms} it was shown that of the four real relative moduli of the SU(2) monopole of charge 
$\bm{k}=(2)$, there was a two dimensional geodesic submanifold corresponding to varying the two moduli 
present in the spectral curve.  This justified the deduction of one dimensional submanifolds in section 
\ref{charge2symmetries}.  The SU(3) monopole of charge $\bm{k}=(1,1)$ also has four real relative moduli, and 
we will assume that the two which appear in the spectral curve again provide a geodesic submanifold.
\par
\begin{figure}
\centering
\includegraphics[width=\linewidth]{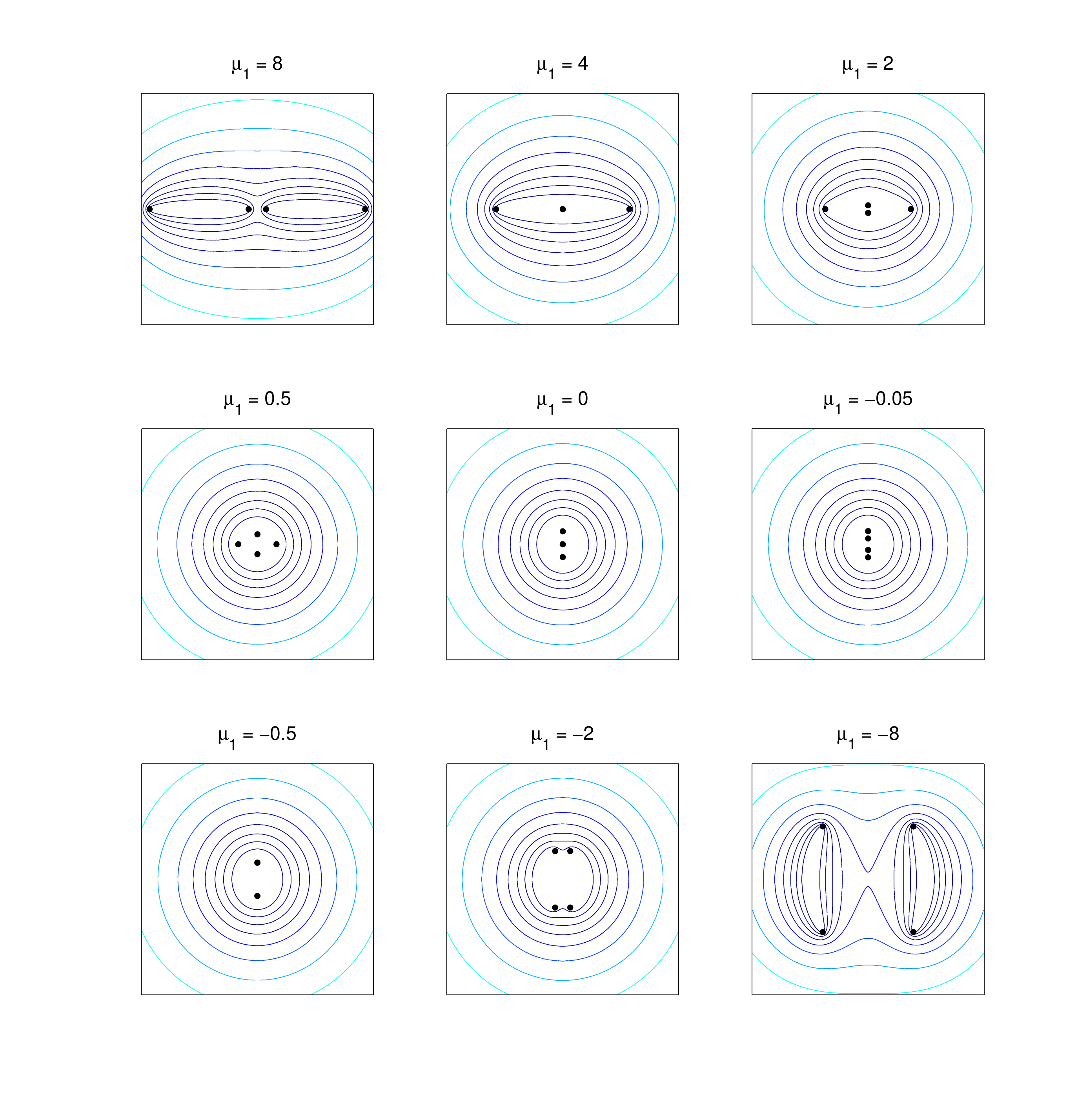}
\caption{Sequence arising by varying the real parameter $\mu_1$ with $\mathfrak{v}_1=0$ and 
$\mathfrak{v}_2=\text{i}\pi$.  Plots show the discriminant of $\hat{\Phi}$.  In the Nahm transformed picture, 
motion of the zeroes of $\text{det}(\Phi)$ follows a similar pattern to that shown in 
fig.~\protect\ref{fig08}, though now with the third zero fixed at $s=0$ and the singularity at 
$s=\text{i}\pi/\beta$.  The zeroes are coincident when $\mu_1=4$ and two of them reach the singularity when 
$\mu_1=0$.}\label{fig16}
\end{figure}

 The reduced moduli are constrained by looking for configurations invariant under a reflection in the 
$x$-axis, which we perform by mapping $\zeta\mapsto\bar{\zeta}$ and $w\mapsto\bar{w}$.  This requires all the 
coefficients $a_{i,j}$ in \eqref{k11coeffs} to be real.  A symmetric choice of boundary conditions is 
provided by requiring the two fundamental monopoles to be of the same size, which we do by further imposing 
invariance of the spectral curve under $\zeta\mapsto-\zeta$ and $w\mapsto w^{-1}$, resulting in
\begin{equation*}
a_{1,1}\,=\,-\text{e}^{\mathfrak{v}_1}\qquad a_{1,0}\,=\,1-\mu_1\text{e}^{\mathfrak{v}_1}\qquad a_{2,1}\,=\,-\text{e}^{\mathfrak{v}_1}\qquad a_{2,0}\,=\,\mu_1\text{e}^{\mathfrak{v}_1}-1,
\end{equation*}
where $\mathfrak{v}_1$ is fixed and $\mu_1\in\mathbb{R}$ provides a one parameter family (note this is a 
different situation from that of section~\ref{trivialembedding}, where $\mathfrak{v}_2=0$).  Fig.~\ref{fig16}
illustrates the resulting scattering process.  As mentioned in \cite{MS07}, the monopoles scatter back off 
each other in a head-on collision, though with a deformed shape.  By allowing different boundary conditions, 
one can in fact find one parameter families describing the less symmetric cases where one monopole is larger 
than the other, or when one of the incoming monopoles is rotated by an angle of $\pi/2$.  As was noted for 
the SU(2) periodic monopole in sections \ref{ch1en} and \ref{charge2symmetries}, we find that when the 
spectral points are well separated those of each fundamental monopole are joined by lines of zero 
discriminant.
\section{Dirac Singularities and the Doubly Periodic Instanton}\label{singularities}
In the region where $z$ dependence can be ignored, the fields of a configuration of positive and negative 
Dirac monopoles at $\zeta=\zeta_i^\pm$ are
\begin{equation*}
\text{i}\beta\hat{\Phi}\,=\,\frac{1}{2}\sum_{i=1}^{n_+}\log\left(|\zeta-\zeta_i^+|^2\right)-\frac{1}{2}\sum_{i=1}^{n_-}\log\left(|\zeta-\zeta_i^-|^2\right)
\end{equation*}
\begin{equation*}
\text{i}\beta\hat{A}_z\,=\,\frac{\text{i}}{2}\sum_{i=1}^{n_+}\log\left(\frac{\bar{\zeta}-\bar{\zeta}_i^+}{\zeta-\zeta_i^+}\right)-\frac{\text{i}}{2}\sum_{i=1}^{n_-}\log\left(\frac{\bar{\zeta}-\bar{\zeta}_i^-}{\zeta-\zeta_i^-}\right)
\end{equation*}
allowing us to compute the holonomy and hence write down the spectral curve,
\begin{equation}
\left(\prod_{i=1}^{n_-}(\zeta-\zeta_i^-)\right)\,w-\left(\prod_{i=1}^{n_+}(\zeta-\zeta_i^+)\right)\,=\,0,\label{u1spec}
\end{equation}
and there are thus no moduli.  Cherkis \& Kapustin \cite{CK03} argue that singularities can be introduced to 
the periodic monopole by modifying the spectral curve \eqref{msc} to
\begin{equation*}
P_{0,n_-}(\zeta)w^N+P_{1,k_1}(\zeta)w^{N-1}+\ldots+P_{N-1,k_{N-1}}(\zeta)w+(-1)^NP_{N,n_+}(\zeta)\,=\,0
\end{equation*}
where $P_{0,n_-}(\zeta)$ and $P_{N,n_+}(\zeta)$ are the monic polynomials appearing in \eqref{u1spec}.
\par The principal use of Dirac singularities is in changing the boundary conditions on the Hitchin data.  In 
particular, adding $K$ positive and $K$ negative singularities to the monopole with $\bm{k}=(K,K,\ldots,K)$ 
renders $\text{det}(\Phi)$ smooth at $|r|\to\infty$, albeit with singularities at finite $|r|$ due to $K$ 
appearing more than once.  We illustrate this by looking at the SU(2) monopole with two singularities, where 
we require the spectral curve to be invariant under $w\mapsto w^{-1}$ in order that the monopole fields are 
valued in $\mathfrak{su}(2)$.  The relevant spectral curve is
\begin{equation}
(\zeta-\zeta_0)w^2-2(a\zeta+b)w+(\zeta-\zeta_0)\,=\,0\label{specsingular}
\end{equation}
such that the boundary conditions \eqref{asympholo} translate to
\begin{equation*}
a\,=\,\cosh(\mathfrak{v})\qquad\qquad b\,=\,\mu\sinh(\mathfrak{v})-\zeta_0\cosh(\mathfrak{v})
\end{equation*}
and the spectral curve \eqref{specsingular} can be rearranged to give the Hitchin Higgs field
\begin{equation}
\Phi\,=\,\zeta\,=\,\zeta_0+\frac{\mu\sinh(\mathfrak{v})}{\cosh(\beta s)-\cosh(\mathfrak{v})}.\label{phising}
\end{equation}
Applying the method of section \ref{spap}, spectral points are located at
\begin{equation*}
\zeta\,=\,\frac{\zeta_0+b}{1-a}\qquad\text{and}\qquad\zeta\,=\,\frac{\zeta_0-b}{1+a},
\end{equation*}
which are centered if $ab+\zeta_0=0$, and are coincident if $a\zeta_0+b=0$.  The monopole Higgs field is
\begin{equation*}
\hat{\Phi}\,=\,\frac{\text{i}}{\beta}\,\Re\cosh^{-1}\left(\cosh(\mathfrak{v})+\frac{\mu\sinh(\mathfrak{v})}{\zeta-\zeta_0}\right)\sigma_3.
\end{equation*}
In the case where $a=0$, $\text{i}b=C$, this simplifies to
\begin{equation*}
\hat{\Phi}\,=\,\frac{\text{i}}{\beta}\,\Re\cosh^{-1}\left(\frac{C}{\zeta}\right)\sigma_3
\end{equation*}
which is related to the fields of sections \ref{spcu} and \ref{ch1en} by a simple inversion transformation 
$\zeta\mapsto C^2/\bar{\zeta}$, with a corresponding change of boundary conditions.
\par In analogy with monopoles appearing as constituents of periodic instantons (see, for example, 
\cite{KvB98a,KvB98b,LL98}), it is expected that the doubly periodic instanton will be related to the periodic 
monopole \cite{CK03,FP+}.  The Nahm data for the doubly periodic instanton are Hitchin equations on a 2-torus 
$T^2$.  The charge 1 case is considered by \cite{FP+}, where the Hitchin system is Abelian.  This allows the 
Hitchin gauge potentials to be expressed as derivatives of a harmonic potential, and the Higgs field is 
chosen to be proportional to $A_s$ in order to share the same singularities,
\begin{equation*}
A_s\,=\,\partial_s\varphi\qquad\qquad A_{\bar{s}}\,=\,-\partial_{\bar{s}}\varphi\qquad\qquad\Phi\,=\,\zeta_0+\alpha\partial_s\varphi
\end{equation*}
where, in our notation, the fundamental solution to Laplace's equation on the torus is
\begin{equation*}
\varphi\,=\,\frac{1}{2}\,\log\frac{\left|\vartheta_3\left(\tfrac{\text{i}}{2\pi}(\bar{s}\beta_1+\bar{\mathfrak{v}})+\tfrac{1}{2}+\tfrac{\text{i}\beta_1}{2\beta_2},\tfrac{\text{i}\beta_1}{\beta_2}\right)\right|^2}{\left|\vartheta_3\left(\tfrac{\text{i}}{2\pi}(\bar{s}\beta_1-\bar{\mathfrak{v}})+\tfrac{1}{2}+\tfrac{\text{i}\beta_1}{2\beta_2},\tfrac{\text{i}\beta_1}{\beta_2}\right)\right|^2}
\end{equation*}
with $\beta_1$ and $\beta_2$ the periods of the instanton, and $\vartheta_3$ is the doubly periodic Jacobi 
theta-function, which is conveniently expressed as
\begin{equation}
\vartheta_3(w,\tau)\,=\,\sum_{n=-\infty}^{\infty}\text{e}^{\text{i}\pi n^2\tau+2\text{i}\pi nw}.\label{jacobitheta}
\end{equation}
The result \eqref{phising} is recovered in the limit $\beta_1=\beta$, $\beta_2\to0$, such that only the $n=0$ 
and $n=-1$ terms contribute to \eqref{jacobitheta},
\begin{equation*}
\varphi\,=\,\frac{1}{2}\,\log\frac{\left|1-\text{e}^{\beta\bar{s}+\bar{\mathfrak{v}}}\right|^2}{\left|1-\text{e}^{\beta\bar{s}-\bar{\mathfrak{v}}}\right|^2}\qquad\Rightarrow\qquad\Phi\,=\,\zeta_0-\frac{\alpha\beta}{2}\frac{\sinh(\mathfrak{v})}{\cosh(\beta s)-\cosh(\mathfrak{v})},
\end{equation*}
which is precisely of the form \eqref{phising}.  In \cite{FP+}, $\alpha$ is interpreted as a size, which when 
set to zero provides axially symmetric fields.  In the monopole picture this corresponds to setting $\mu=0$, 
in which case $a\zeta_0+b=0$ and the spectral points coincide, again leading to axial symmetry.
\par The need for singularities when making the comparison with the doubly periodic instanton is brought 
about by a change in the boundary conditions, and is reminiscent of the intepretation of periodic instantons 
as monopoles whose gauge group is a loop group \cite{Har08}.  In practice, this amounts to adding a root to 
the gauge group such that all of the $\ell_i$ vanish and we are at the origin of the root diagram, 
fig.~\ref{fig01}.  From the discussion of sections \ref{introduction} and \ref{spap}, the additional 
fundamental monopole expected from the extra root fits in with the observation in \cite{FP+} that the doubly 
periodic instanton consists of two periodic monopole constituents, separated in one of the periodic 
directions.  It would be interesting to explore this result further, although this would require a departure 
from the approximation presented in this paper.
\section{Concluding Remarks}\label{outlook}
In this paper we developed a technique, motivated by \cite{CK01,CK03,War05,HW09}, to study the singly 
periodic BPS monopole.  This was checked against numerical studies of the SU(2) cases of charge $1$ and $2$.  
Geodesic motion on an effective two dimensional moduli space compared favourably with analytic results for 
charge $2$.  In particular, it was found that motion transverse to the periodic direction provides a geodesic 
submanifold.  Some simple SU(3) configurations and singular periodic monopoles were also considered in this 
context.  The Nahm transform relates the periodic monopole to a Hitchin system on the cylinder, giving rise 
to lumps whose motion is described, at large separations, by the motion of zeroes of the spectral curve 
polynomial.
\par Short of finding explicit solutions for the monopole fields or the moduli space metric, some unanswered 
questions which will provide the basis for future work include whether the two energy peaks associated to 
each fundamental monopole can be understood as `constituents' in their own right.  This has been done for the 
periodic instanton, which was reconstructed by \cite{LL98} in terms of the Nahm data of its monopole 
constituents.  It would also be of interest to study explicitly the limits of the periodic monopole for large 
and small periods.  Preliminary numerical work indicates the Nahm data behaves as hoped (see section 
\ref{lumpscyl}).  As a step in this direction, a study of how the moduli describing a phase difference and 
$z$ separation appear in the Nahm dual picture will be presented in \cite{MW}.
\section*{Acknowledgements}
The author wishes to thank R. S. Ward for guidance and James P. Allen and Derek Harland for useful 
discussions.  This research has been supported by an STFC studentship.
\appendix
\section{Symmetries of the Nahm Operator}\label{appen}
In this appendix we explain in detail the procedure followed in section \ref{fullsymms}, with reference to 
the example of the $K\in\mathbb{R}$ geodesic of the `zeroes together' solution.
\par The map $(s;K)\mapsto(\bar{s};\bar{K})$ transforms $(r,t)\mapsto(r,-t)$,
\begin{equation*}
\mu_+(s;K)\,=C\cosh(\beta s)+K/2\,\mapsto\,C\cosh(\beta\bar{s})+\bar{K}/2\,=\,\bar{\mu}_+(s;K)
\end{equation*}
and
\begin{equation*}
\mu_-(s;K)\,=\,1\,\mapsto\,\bar{\mu}_-(s;K).
\end{equation*}
Equation \eqref{psieq} is invariant, so $\Re(\psi)(s;K)\mapsto\Re(\psi)(s;K)$.  Recalling that in this 
case $\Im(\psi)=0$ gives the transformation of $a$,
\begin{equation*}
a(s;K)\,=\,-\tfrac{1}{8}(\partial_r+\text{i}\partial_t)\psi\,\mapsto\,-\tfrac{1}{8}(\partial_r-\text{i}\partial_t)\psi\,=\,\bar{a}(s;K).
\end{equation*}
Combining these results we obtain the transformed Hitchin fields \eqref{pg},
\begin{equation*}
\Phi(s;K)\,=\begin{pmatrix}0&\mu_+\text{e}^{\psi/2}\\\mu_-\text{e}^{-\psi/2}&0\end{pmatrix}(s;K)\mapsto\,\Phi'(s';K')\,=\,\begin{pmatrix}0&\bar{\mu}_+\text{e}^{\psi/2}\\\bar{\mu}_-\text{e}^{-\psi/2}&0\end{pmatrix}(s;K),
\end{equation*}
\begin{equation*}
A_{\bar{s}}(s;K)\,=\,a(s;K)\sigma_3\,\mapsto\,A_{\bar{s}}'(s';K')\,=\,\bar{a}(s;K)\sigma_3\,=\,-A_s(s;K)
\end{equation*}
\begin{equation*}
A_s(s;K)\,=\,-\bar{a}(s;K)\sigma_3\,\mapsto\,A_s'(s';K')\,=\,-a(s;K)\sigma_3\,=\,-A_{\bar{s}}(s;K).
\end{equation*}
The Nahm operator $\Delta$ constructed from the new fields is
\begin{equation*}
\Delta'\,=\,\begin{pmatrix}\bm{1}_2\otimes(2\partial_s-z)-2A_s&\bm{1}_2\otimes\zeta-(\Phi')^\dag\\\bm{1}_2\otimes\bar{\zeta}-\Phi'&\bm{1}_2\otimes(2\partial_{\bar{s}}+z)-2A_{\bar{s}}\end{pmatrix}.
\end{equation*}
Noting that $\Phi'$ can be written in terms of $\Phi$ as $\Phi'=\sigma_1\Phi^\dag\sigma_1$, the new Nahm 
operator $\Delta'$ can be obtained from the original one \eqref{nahmopcharge2} by the combined transformation
\begin{equation*}
\Delta'\,=\,U^{-1}\Delta U\qquad\qquad(\zeta,z)\,\mapsto\,(\bar{\zeta},-z)
\end{equation*}
with $U=\sigma_1\otimes\sigma_1$.  Consequently, $\Psi$ transforms as
\begin{equation*}
\Psi_\pm\,\mapsto\,\sigma_1\Psi_\mp
\end{equation*}
such that the new monopole fields evaluated at $(\bar{\zeta},-z)$ are the same as the old ones at 
$(\zeta,z)$.  A monopole configuration symmetric under $(\zeta,z)\mapsto(\bar{\zeta},-z)$ is thus invariant 
under $K\mapsto\bar{K}$, and leaves us with the one parameter family of solutions described by $\Im(K)=0$.

\end{document}